\documentclass[useAMS,usenatbib,usegraphicx]{mn2e}

\usepackage{epsfig,url}

\def\spose#1{\hbox to 0pt{#1\hss}}
\def\lta{\mathrel{\spose{\lower 3pt\hbox{$\mathchar"218$}}
     \raise 2.0pt\hbox{$\mathchar"13C$}}}
\def\gta{\mathrel{\spose{\lower 3pt\hbox{$\mathchar"218$}}
     \raise 2.0pt\hbox{$\mathchar"13E$}}}

\title[Stability in the $\nu$ Octantis system]{The stability of the suggested planet
in the $\nu$ Octantis system: a numerical and statistical study}
\author[B. Quarles, M. Cuntz, and Z. E. Musielak]{B. Quarles,$^{1}$\thanks{E-mail:
billyq@uta.edu (BQ); cuntz@uta.edu (MC); zmusielak@uta.edu (ZEM)} M. Cuntz,$^{1}$ and Z. E. Musielak$^{1}$\\
$^{1}$University of Texas at Arlington, Department of Physics, 502 Yates Street, Arlington, TX 76019, USA\\}

\begin{document}

\date{Accepted ... Received ...; in original form ...}

\pagerange{\pageref{firstpage}--\pageref{lastpage}} \pubyear{2011}

\maketitle

\label{firstpage}

\begin{abstract}
 We provide a detailed theoretical study aimed at the observational
 finding about the $\nu$~Octantis binary system that indicates the
 possible existence of a Jupiter-type planet in this system.  If
 a prograde planetary orbit is assumed, it has earlier been argued
 that the planet, if existing, should be located outside the zone
 of orbital stability.  However, a previous study by Eberle \& Cuntz
 (2010) [ApJ 721, L168] concludes that the planet is most likely stable
 if assumed to be in a retrograde orbit with respect to the secondary
 system component.  In the present work, we significantly augment
 this study by taking into account the observationally deduced
 uncertainty ranges of the orbital parameters for the stellar
 components and the suggested planet.  Furthermore, our study
 employs additional mathematical methods, which include
 monitoring the Jacobi constant, the zero velocity function, and the
 maximum Lyapunov exponent.  We again find that the suggested planet
 is indeed possible if assumed to be in a retrograde orbit, but it is
 virtually impossible if assumed in a prograde orbit.  Its existence
 is found to be consistent with the deduced system parameters
 of the binary components and of the suggested planet, including the
 associated uncertainty bars given by observations.
\end{abstract}


\section{Introduction}

The binary system $\nu$~Octantis, consisting of a K1~III star
\citep{hou75} and a faint secondary component, identified
as spectral type K7--M1~V \citep{ram09} is located in the
southernmost portion of the southern hemisphere, i.e., in
close proximity to the Southern Celestial Pole.  In 2009,
\citeauthor{ram09} found new detailed astrometric--spectroscopic
orbital solutions, implying the possible existence of
a planet\footnote{An alternative explanation involves
a hierarchical triple system, invoking a precessional motion
of the main stellar component, as proposed by \cite{mor12}.}
around the main stellar component.  Thus, $\nu$~Octantis~A
is one of the very few cases where a planet is found to orbit
a giant rather than a late-type main-sequence star; the first case
of this kind is the planetary-mass companion to the G9~III star
HD~104985 \citep{sat03}.

The composition of the $\nu$~Octantis system leads to many
open questions particularly about the orbital stability of the 
proposed planet.  The semimajor axis of the binary system has
been deduced as $2.55 \pm 0.13$ AU with an eccentricity given as
$0.2358 \pm 0.0003$ \citep{ram09}.  Furthermore, the semimajor
axis of the planet, with an estimated minimum mass $M_p {\sin}i$
of approximately 2.5~$M_J$, was derived as $1.2 \pm 0.1$ AU (see 
Table~\ref{tab:Parameters}). 
The distance parameters indicate that the proposed 
planet is expected be located about halfway between the two 
stellar components. This is a set-up in stark contrast to other observed 
binary systems hosting one or more planets \citep{egg04,rag06}
and, moreover, difficult to reconcile with standard orbital
stability scenarios.

A possible solution was proposed by
\cite{ebe10b}. They argued that the planet might most likely be 
stable if assumed to be in a retrograde orbit relative to the
motion of the secondary stellar component.  Previous cases
of planetary retrograde motion have been discovered, but with respect to the
spin direction of the host stars.  Examples include (most likely)
HAT-P-7b (or Kepler-2b), WASP-8b and WASP-17b that were identified
by \cite{win09}, \cite{que10}, and \cite{and10}, respectively.
These discoveries provide significant challenges to the
standard paradigm of planet formation in protoplanetary disks
as discussed by \cite{lis93}, \cite{lin96} and others.

The previous theoretical work by \cite{ebe10b} explored two
cases of orbital stability behaviour for $\nu$~Octantis based
on a fixed mass ratio for the stellar components and a starting
distance ratio of $\rho_0 = 0.379$ for the putative planet (see
definition below).  They found that if the planet is placed in
a prograde orbit about the stellar primary, it almost immediately
faced orbital instability, as expected.  However, if placed in
a retrograde orbit, stability is encountered for at least
$1 \times 10^7$~yr or $3.4 \times 10^6$ binary orbits, i.e.,
the total time of simulation.  This outcome is also consistent
with the results from an earlier study by \cite{jef74} based on
a Henon stability analysis, which indicates significantly enlarged
regions of stability for planets in retrograde orbits.  

Nonetheless, there is a significant need for augmenting the
earlier study by \cite{ebe10b}.  Notable reasons include:
first, \citeauthor{ebe10b} only considered one possible mass
ratio and one possible set of values for the semimajor axis and
eccentricity for the stellar components, thus disregarding their
inherent observational uncertainties.  Hence, it is unclear
if or how the prograde and retrograde stability regions for the
planetary motions will be affected if other acceptable choices
of system parameters are made.  Second, \citeauthor{ebe10b}
chose the 9 o'clock position as the starting position for the
planetary component.  Thus, it is unclear if or how the timescale
of orbital stability will change due to other positional choices
in the view of previous studies, which have demonstrated a significant
sensitivity in the outcome of stability simulations on the
adopted planetary starting angle \citep[e.g.,][]{fat06,yea11}.
Third, and foremost, \citeauthor{ebe10b} did not utilize 
detailed stability criteria for the identification of 
the onset of orbital instability.  Possible mathematical methods 
for treating the latter include monitoring the Jacobi constant and 
the zero velocity function \citep[e.g.,][]{sze67,ebe08} as well as
using the maximum Lyapunov exponent \citep[e.g.,][]{hil94,smi93,
lis99,mur01}.  In our previous work reported by \cite{cun07},
\cite{ebe08} and  \cite{qua11}, we extensively used these methods
for investigating the orbital stability in the circular restricted 3-body
problem.  The same methods will be adopted in the present study.

Our paper is structured as follows.  In Sect.~2, we discuss
our theoretical methods, including the initial conditions of
the star-planet system and the stability criteria.  The results
and discussion, including detailed case studies, are given in
Sect.~3.  Finally, Sect.~4 conveys our conclusions.

\begin{table*}
\begin{center}
\caption{Stellar and Planetary Parameters of $\nu$ Octantis}
\vspace{0.05in}
\vspace{0.05in}
\begin{tabular}{l c c }
\hline
\hline
\noalign{\vspace{0.03in}}
${\centering  {\rm Parameter}}$ & Value & Reference \\
\noalign{\vspace{0.03in}}
\hline
\noalign{\vspace{0.03in}}
Spectral type (1)                 & K1 III                                            & \cite{hou75} \\
Spectral type (2)                 & K7$-$M1 V                                         & \cite{ram09} \\
R.A.                              & 21$^{\rm h}$ 41$^{\rm m}$ 28.6463$^{\rm s}$       & \cite{esa97}$^{a,b}$ \\
Decl.                             & $-$77$^\circ$ 23$^\prime$ 24.167$^{\prime\prime}$ & \cite{esa97}$^{a,b}$ \\
Distance (pc)                     & 21.20 $\pm$ 0.87                                  & \cite{lee07} \\
$M_v$ (1)                         & 2.10 $\pm$ 0.13                                   & \cite{ram09}$^c$ \\
$M_v$ (2)                         & $\sim$ 9.9                                        & \cite{dri00} \\
$M_1$ $(M_\odot)$                 & 1.4 $\pm$ 0.3                                     & \cite{ram09} \\ 
$M_2$ $(M_\odot)$                 & 0.5 $\pm$ 0.1                                     & \cite{ram09} \\
$T_{\rm eff,1}$ (K)               & 4790 $\pm$ 105                                    & \cite{all99} \\
$R_1$ $(R_\odot)$                 & 5.9 $\pm$ 0.4                                     & \cite{all99} \\
$P_b$ (d)                         & 1050.11 $\pm$ 0.13                                & \cite{ram09} \\
$a_b$ (AU)                        & 2.55 $\pm$ 0.13                                   & \cite{ram09} \\
$e_b$                             & 0.2358 $\pm$ 0.0003                               & \cite{ram09} \\
$M_p{\rm sin} \textit{i}$ $(M_J)$ & 2.5 $\pm$ $\ldots$                                & \cite{ram09}$^d$ \\
$a_p$ (AU)                        & 1.2 $\pm$ 0.1                                     & \cite{ram09}$^d$ \\
$e_p$                             & 0.123 $\pm$ 0.037                                 & \cite{ram09}$^d$ \\
\noalign{\vspace{0.05in}} \hline
\end{tabular}
\label{tab:Parameters}
\vspace{0.05in}
\begin{enumerate}
\item[$^a$] Data from SIMBAD, see \url{http://simabd.u-strasbg.fr}.
\item[$^b$] Adopted from the $\textit{Hipparcos}$ catalog.
\item[$^c$] Derived from the stellar parallax, see \cite{ram09}.
\item[$^d$] Controversial.
\end{enumerate}
\end{center}
\end{table*}

\section{Theoretical Approach}

\subsection{Basic Equations}
\label{sec:BE}

Using the orbital parameters attained by \cite{ram09}, see
Table~\ref{tab:Parameters}, we consider the coplanar restricted
3-body problem (RTP) applied to the system of $\nu$ Octantis.  Following
the standard conventions pertaining to the coplanar RTP, we write
the equations of motion in terms of the parameters $\mu$ and $\rho_0$.
The parameter $\mu$ and the complementary parameter $\alpha$ are defined
by the two stellar masses $m_1$ and $m_2$ (see below).  The planetary
distance ratio $\rho_0$ depends on $R_0$ and $D$, which denote the
initial distance of the planet from its host star, the more massive
of the two stars with mass $m_1$, and the initial distance between the
two stars, respectively.  In addition, we represent the system in a
rotating reference frame, which introduces Coriolis and centrifugal
forces.  The following equations of motion utilize these parameters \citep{sze67}:

\begin{eqnarray}
\dot{x} & = & u \\
\dot{y} & = & v \\
\dot{z} & = & w \\
\dot{u} & = & 2v + \Omega(x -\alpha {x - \mu \over r_{1}^3} -\mu {x + \alpha 
\over r_{2}^3}) \\
\dot{v} & = & -2u + \Omega(y -\alpha {y \over r_{1}^3} -\mu {y \over r_{2}^3}) \\
\dot{w} & = & -z + \Omega(z-\alpha {z \over r_{1}^3} -\mu {z \over r_{2}^3})
\end{eqnarray}
where
\begin{eqnarray}
\mu     & = & {m_{2} \over m_{1} + m_{2}} \\
\alpha  & = & 1 - \mu \\
\Omega  & = & (1 +  e\: {\rm cos} \:f)^{-1} \\
r_{1} & = & \sqrt{\left(x - \mu\right)^2 + y^2 + z^2} \\
r_{2} & = & \sqrt{\left(x + \alpha\right)^2 + y^2 + z^2}\\
\rho_0 & = & {R_0 \over D}
\end{eqnarray}

The variables in the above equations describe how the state of a planet,
commensurate to a test object, evolves due to the forces present.
The state is represented in a Cartesian coordinate system
$\left\{x,\;y,\;z,\;u,\;v,\;w\right\}$.  We use dot notation
to represent the time derivatives of the coordinates
$\left\{\dot{x}={\rm dx \over dt}\right\}$.  Since the variables
$\left\{u,\;v,\;w\right\}$ represent the velocities, the time derivatives
of these variables represent the accelerations.  The variables
$\left\{r_{1},\;r_{2}\right\}$ denote the distances of the planet
from each star in the rotating reference frame.  The value of $f$ describes
the true anomaly of the mass $m_1$ and $e$ refers to the eccentricity of
the stellar orbits.

\subsection{Initial Conditions}
\label{sec:IC}

\begin{figure*}

\centering
\begin{tabular}{cc}
\includegraphics[trim = 1mm 1mm 1mm 1mm, clip, width=0.4\linewidth]{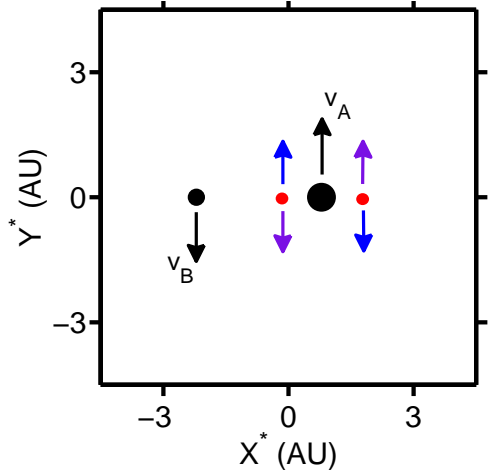}\\
\end{tabular}
\caption{Example of the possible initial starting configurations.
The directions of the initial velocities of the stellar components
($\nu$~Oct~A and $\nu$~Oct~B) are shown with black coloured arrows
and appropriate labels.  The proposed planet is represented by the
red coloured dots in the 3 o'clock position (right) and the
9 o'clock position (left).  The possible directions of the
initial velocity for the planetary component are either
prograde (purple) or retrograde (blue).  The arrows indicate the
directions in the initial configurations of the system.
}
\label{fig:Fig01}
\end{figure*}

The coplanar RTP indicates that the putative planet moves in the same orbital plane
as the two stellar components; additionally, its mass is considered
negligible\footnote{Negligible mass means that although the body's motion
is influenced by the gravity of the two massive primaries, its mass is too
low to affect the motions of the primaries.}.  The initial orbital velocity of
the planet is calculated assuming a Keplerian circular orbit about its host star;
see Fig.~\ref{fig:Fig01} for a depiction of the initial system configuration.
For simplicity we assume the $(u,w)$ components are equal to zero and the
magnitude of the velocity to be in the direction of the $v$ component.  The
mean value of the eccentricity of the binary, $e_b$, is used due to accuracy
of the value given.  A Taylor expansion in initial orbital velocity of the
binary shows that the first order correction would only change the
initial orbital velocity of the respective stars by a factor of $10^{-8}$.
The initial eccentricity of the planet, $e_p$, has been chosen to be equal
to zero due to the redundancy it provides in the selection of the initial
planetary position.  Noting that the inclination angle of the planet, $i_p$,
has not yet been determined, it is taken to be $90^\circ$ consistent with the
requirement of the coplanar RTP.

To integrate the equations of motion, a sixth order symplectic integration
method is used \citep{yos90}.  In the computation of Figs.~\ref{fig:Fig02}
and~\ref{fig:Fig03} a fixed time step of $10^{-3}$ is assumed.  This
proved to be appropriate since the average relative error in the Jacobi constant
is found to be on the order of $10^{-8}$.  To increase the precision for the
individual cases depicted in Figs.~\ref{fig:Fig05} to \ref{fig:Fig09}, a fixed
time step of $10^{-4}$ is used, which decreased the average relative error
to the order of $10^{-10}$.  However, this choice entailed a considerable
increase of the computation time.  Therefore, the models pertaining to
Fig.~\ref{fig:Fig02} were integrated for 10,000 binary orbits (approximately
29,000 years) and the models pertaining to Fig.~\ref{fig:Fig03} were integrated
for 1,000 binary orbits (approximately 2,900 years), noting that this figure
is considerably less than utilized in the mainframe of our study for the
determination of the statistical probability of a prograde planetary configuration.
Moreover, the models of Figs.~\ref{fig:Fig05} to \ref{fig:Fig09} were integrated
for 80,000 binary orbits (approximately 232,000 years). 

In the treatment of this problem, four separate types of initial conditions
of the system are considered.  They encompass the two different starting positions
of the planet with respect to the massive host star, which are: the 3 o'clock
and 9 o'clock positions (see Fig.~\ref{fig:Fig01}).  With each of these starting
positions, we considered
the possibility of planetary motion in counter-clockwise and clockwise direction
relative to the direction of motion of the secondary stellar component; the
directions are referred to as prograde and retrograde, respectively.  The
primary masses are considered to move in ellipses about the centre of mass
of the system.

\subsection{Statistical Parameter Space}
\label{sec:SPS}

In the present work we will explore the role(s) of small variations in the
system parameters. Notably we will investigate how the stellar mass ratio as 
well as the semimajor axis and inherently the eccentricity of the
stellar components can change the outcome of our orbital stability simulations.
Particularly when we consider the statistical probability
of prograde orbits for the suggested planet.  These variations are
due to the observational uncertainties of the respective parameter;
see \cite{ram09} and Table~\ref{tab:Parameters} for details.

Equation (7) gives the definition of the mass ratio $\mu$ with
$m_1 = 1.4 \pm 0.3~M_\odot$ and $m_2 = 0.5 \pm 0.1~M_\odot$.
Taking $m_1$ and $m_2$ as statistically independent from one another,
the bounding values for $\mu$ are $\left[0.220, 0.306\right]$,
i.e., $0.263 \pm 0.043$.  In this case, we assumed the reported
uncertainty bars for the stellar masses to be $1.25~\sigma$, noting
that through this choice $\mu$ is consistent with the mass ratio 
$q = 0.38 \pm 0.03$ derived by \cite{ram09} that was used for computing
$m_2$ based on the sum of the two masses.  Formulas and methods for
the propagation-of-error, which in essence employ a first-order
Taylor expansion, have been given by, e.g., \cite{mey75}; see also
\cite{pre86} for numerical methods if the error is calculated via
a Monte-Carlo simulation.

Equation (12) gives the definition for the relative initial distance
of the planet, expressed in terms of $R_0$ and $D$, i.e., the initial
distance of the planet from its host star, the more massive of the two
stars with mass $m_1$ and the initial distance between the two stars,
respectively.  This definition also allows to compute the bounding
values for $\rho_0$.  Following the assumption made by \cite{ebe10b},
the set up of the binary system will be initialized by assuming the
secondary stellar component to be located in the apastron starting position,
entailing that $D$ is modified as $D = a_b\left(1+e_b\right)$.  Thus, the
observation uncertainties of $a_p$, $a_b$, and $e_b$ will be utilized
to compute the bounding values for $\rho_0$ (see Table~\ref{tab:Parameters}).
As result we obtain $\left[0.344, 0.418\right]$, i.e., $0.381 \pm 0.037$.
These values will be considered in the
subsequent set of detailed orbital stability studies by making
appropriate choices for $\mu$ and $\rho_0$ when setting the initial
conditions.  The outcome of the various simulations will, in particular,
allow a detailed statistical analysis of the mathematical probability
for the existence of the suggested planet in a prograde orbit; see
Sect.~\ref{sec:StatS}.

\subsection{Stability Criteria}
\label{sec:Criteria}

In our approach we employ two independent criteria.  One method
deals with identifying events that can cause instability, whereas
the other method allows us to determine stability.  First, we monitor
the relative error in the Jacobi constant to detect instabilities.
By monitoring this constant we can determine when the integrator
loses its accuracy or fails completely.  It has been shown that this
method is particularly sensitive to close approach events of the planet
with any of the stellar components.  However, it has previously been
shown by \cite{ebe08} and \cite{ebe10a} that these close approach events
are preceded by an encounter with the so-called ``zero velocity contour"
(ZVC) for the coplanar circular RTP.

Generalizing this criteria to the elliptic RTP, an analogous ``zero velocity
function" (ZVF) can be found that is, however, dependent on the initial
distance between the stellar components \citep{sze67}.  The ZVF can also
be described by a pulsating ZVC; see \cite{sze08} for updated results.
Note that the ZVC will be largest when the binary components are at apastron
and smallest when they are at periastron.  Noting that we choose the stars
to be at the apastron starting position, the largest ZVC is inherently defined
as well.  When the planetary body encounters the ZVC, the orbital velocity
decreases dramatically, which reduces the $\dot{u}$ and $\dot{v}$-related
forces.  As a result, the planetary body follows a trajectory that results
in a close approach.  This behaviour provides an indication that the system
will eventually become unstable, resulting in an ejection of the planetary
body from the system (most likely outcome) or its collision with one of the
stellar components (less likely outcome).

The second criterion utilizes the method of Lyapunov exponents \citep{lya07},
which is commonly used in nonlinear dynamics for determining the onset of chaos
in different dynamical systems \citep[e.g.,][]{hil94,mus09}.  A numerical 
algorithm that is typically adopted for calculating the spectrum of Lyapunov exponents 
was originally developed by \cite{wol85}.  In this approach, a set of tangent 
vectors $\textbf{x}_{i}$ and their derivatives $\dot{\textbf{x}}_{i}$ with 
$i=1,\ldots,6$ are introduced and initialized. In our approach, we choose 
all tangent vectors to be unit vectors for simplicity, and use a standard 
integrator akin to the Runge-Kutta schemes for computing changes in the tangent 
vectors within each timestep.  After the tangent vectors become oriented along 
the flow, it is necessary to perform a Gram-Schmidt Renormalization (GSR) to 
orthogonalize the tangent space.  In this specific procedure, we calculate
a new orthonormal set of tangent vectors here denoted by primes ($\prime$), 
and obtain

\begin{eqnarray}
\textbf{x}_1^{'} & = & {\textbf{x}_1 \over \left\|\textbf{x}_{1}\right\|} \\
\textbf{x}_2^{'} & = & {\textbf{x}_2 - \left\langle\textbf{x}_2,\:\textbf{x}_1^{'} \right\rangle 
\textbf{x}_1^{'} \over \left\|\textbf{x}_2 - \left\langle\textbf{x}_2,\:\textbf{x}_1^{'} 
\right\rangle \textbf{x}_1^{'}\right\|} \\
\vdots \nonumber \\
\textbf{x}_6^{'} & = & {\textbf{x}_6 - \left\langle\textbf{x}_6,\:\textbf{x}_5^{'} \right\rangle 
\textbf{x}_5^{'} - \ldots -\left\langle\textbf{x}_6,\:\textbf{x}_1^{'} \right\rangle 
\textbf{x}_1^{'} \over \left\|\textbf{x}_6 - \left\langle\textbf{x}_6,\:\textbf{x}_5^{'} 
\right\rangle \textbf{x}_5^{'} - \ldots -\left\langle\textbf{x}_6,\:\textbf{x}_1^{'} 
\right\rangle \textbf{x}_1^{'}\right\|}
\end{eqnarray}

With this new set of tangent vectors, the spectrum of Lyapunov exponents 
is computed by using

\begin{equation}
\lambda_i \ = \ {1\over \tau}\left[\lambda_{i-1} + \log\left\|\textbf{x}_i - \sum_{j=1}^{i}
{\left\langle \textbf{x}_i,\:\textbf{x}_{j-1}^{'}\right\rangle\textbf{x}_{j-1}^{'}}\right\| 
\right]
\end{equation}
where $\lambda_{o}=\textbf{x}_{o}^{'}=0$.

The Lyapunov exponents have extensively been used in studies of orbital stability 
in different settings pertaining to the Solar System \citep[e.g.,][]{lec92,mil92,lis99,
mur01}.  Typically, the maximum Lyapunov exponent (MLE) is adopted as a measure of 
the rate with which nearby trajectories diverge.  The reason is that the MLE 
offers the unique measure of the largest divergence.  We previously used the MLE criterion 
to provide a cutoff for which stability can be assured given the behaviour of the MLE 
in time.  This cutoff value has previously been determined by \cite{qua11}; it is 
given as log~$\lambda_{\rm max}=-0.82$ on a logarithmic scale of base 10.

Recalling that
the cutoff in MLE is inversely related to the Lyapunov time $t_{\rm L}$, the latter
provides an estimate of the predictive period for which stability is guaranteed.
The lower (i.e., more negative) the logarithmic value of the MLE, the longer the
time for which orbital stability will be ensured.  Due to the finite time of integration,
this estimate is only valid for the integration timescale considered.  But the nature of
how the MLE changes in time will provide the evidence to whether the MLE is asymptotically
decreasing and thereby allowing to identify orbital stability.  Hence this cutoff will
be used for establishing orbital stability of the suggested planet in the $\nu$~Octantis
system.  

\begin{table*}

\begin{center}
\caption{Survival Counts of the Suggested Planet in $\nu$ Octantis for 1,000 binary orbits.}
\vspace{0.05in}
\vspace{0.05in}
\begin{tabular}{l c c c c }
\hline
\hline
\noalign{\vspace{0.03in}}
\multicolumn{2}{c}{Configuration} & Unstable Count & Stable Count & Stable Percentage  \\
\noalign{\vspace{0.03in}}
\hline
\noalign{\vspace{0.03in}}
Prograde   &   3 o'clock  & 17544  &  1776  &  9.2~\%  \\
Prograde   &   9 o'clock  &  19200  &  120  &  0.6~\%  \\
Retrograde &   3 o'clock  & 1040  &  19184  & 94.9~\%  \\
Retrograde &   9 o'clock  &    0  &  20224  & 100~\%   \\
\noalign{\vspace{0.05in}} \hline
\end{tabular}
\label{tab:SC}
\end{center}

\end{table*}

\section{Results and Discussion}

\subsection{Model Simulations}
\label{sec:ModelS}

\begin{figure*}

\centering
\begin{tabular}{cc}
\includegraphics[trim = 3mm 1mm 1mm 1mm, clip, width=0.4\linewidth]{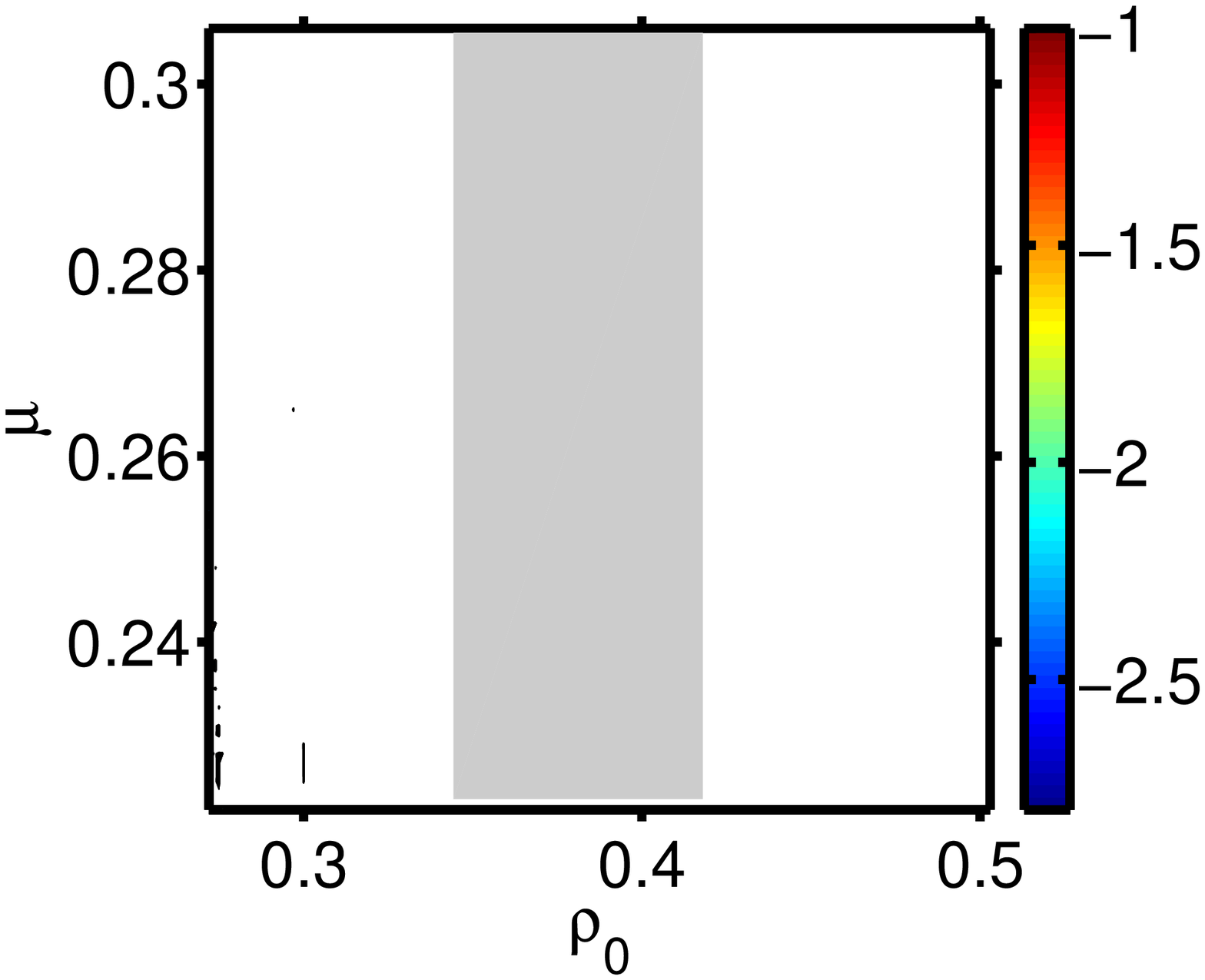}&
\includegraphics[trim = 3mm 1mm 1mm 1mm, clip, width=0.4\linewidth]{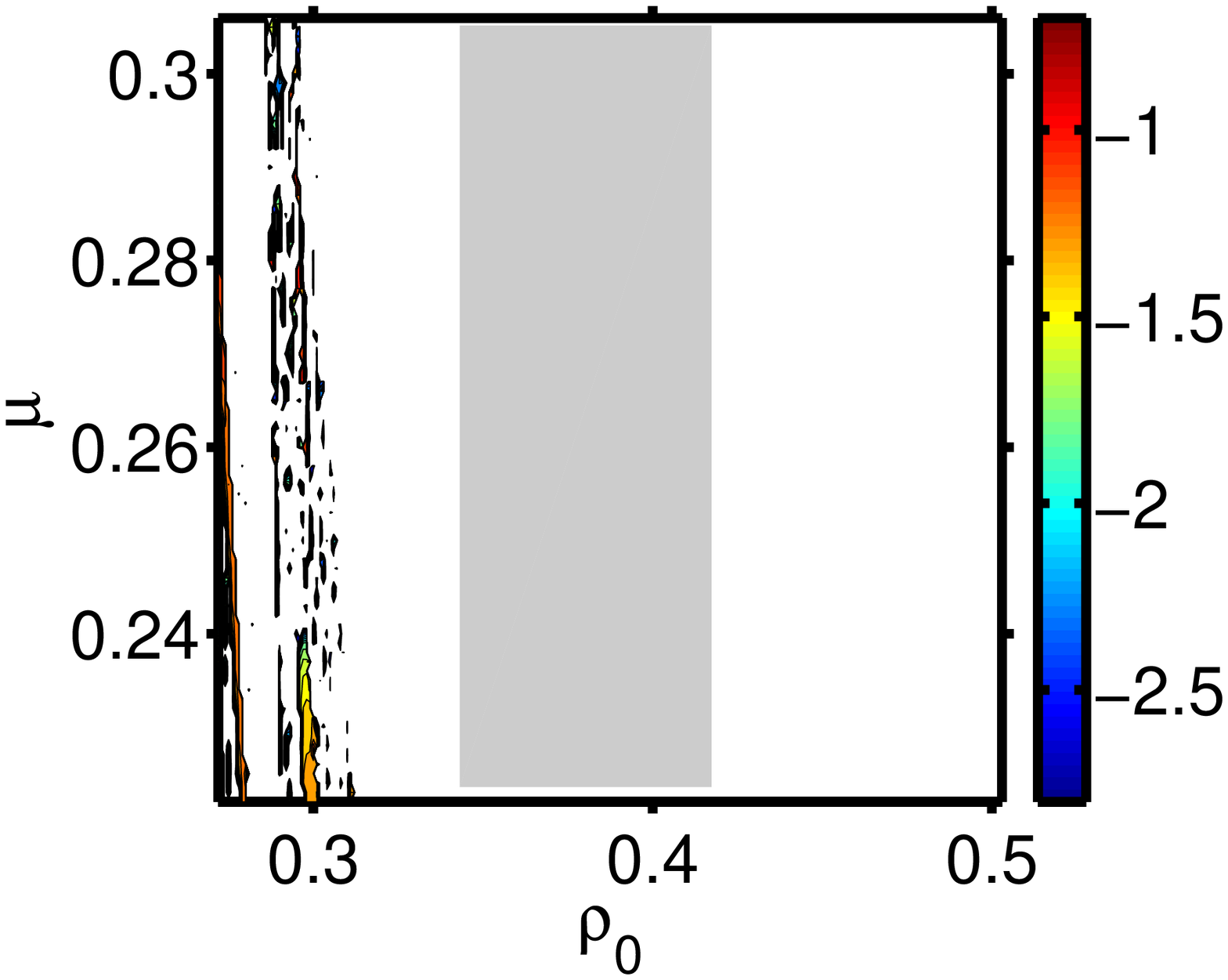}\\
\end{tabular}
\caption{Detailed results of simulations for the parameter space for the case of prograde planetary
motion.  The left and right panel represent the simulations with the planet initially placed at the
9 o'clock and 3 o'clock position, respectively.  The grey area represents the statistical $1 \sigma$
range in $\rho_0$, which is $0.381 \pm 0.037$.
}
\label{fig:Fig02}
\end{figure*}

\begin{figure*}

\centering
\begin{tabular}{cc}
\includegraphics[trim = 0mm 1mm 1mm 1mm, clip, width=0.4\linewidth]{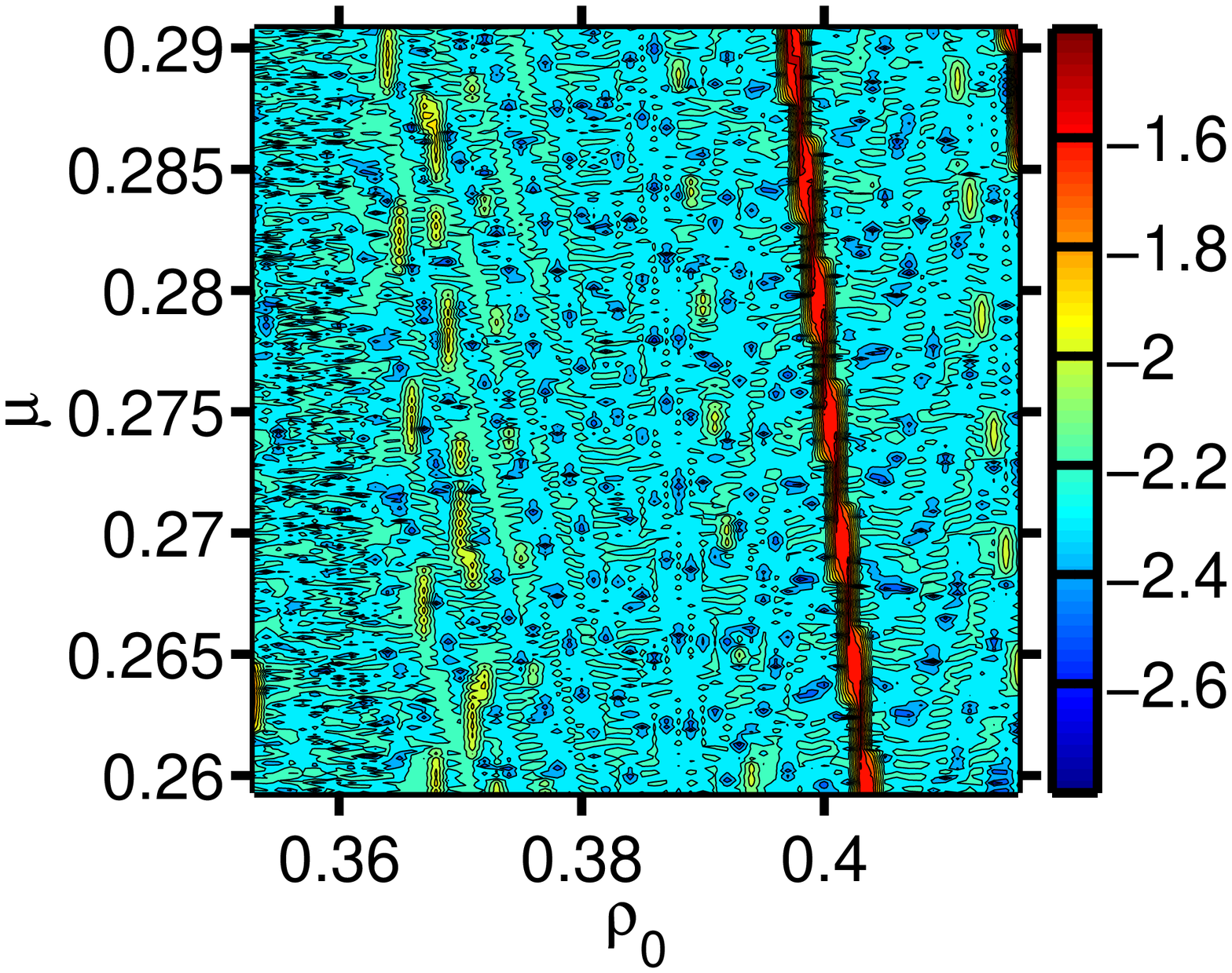}&
\includegraphics[trim = 0mm 1mm 1mm 1mm, clip, width=0.4\linewidth]{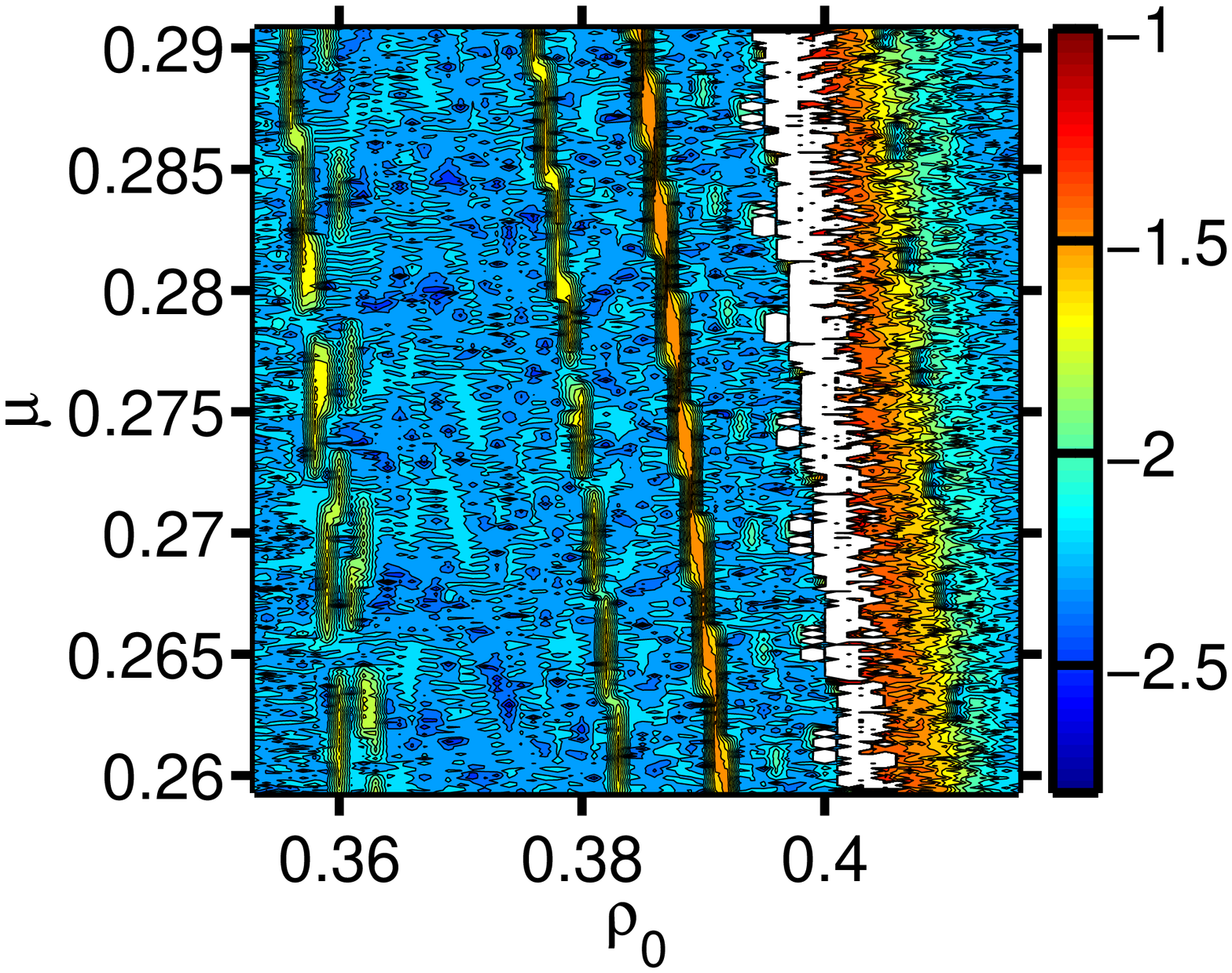}\\
\end{tabular}
\caption{Same as Fig.~\ref{fig:Fig02}, but now for retrograde planetary motion.}
\label{fig:Fig03}
\end{figure*}

In the following we report the general behaviour of stability for the set of
orbital stability simulations pursued in our study.  Our initial focus
is on the MLE.  In Figs.~\ref{fig:Fig02} and~\ref{fig:Fig03}, the various values
of the MLE as a function of $\mu$ and $\rho_0$ are denoted using a colour code.
The range of colours shows that dark red represents a relatively high
(less negative) logarithmic MLE value, whereas dark blue represents a
relatively low (more negative) logarithmic MLE value.
The white background represents conditions, which failed to complete the respective
allotted run time; see Sect.~\ref{sec:Criteria} for a discussion of possible
conditions for terminating a simulation.

Figure~\ref{fig:Fig02} shows the result of our simulations for 10,000 binary
orbits, i.e., about 29,000 years, in the prograde configuration for
starting conditions given by the 3 o'clock or 9 o'clock planetary position.
The results were obtained using values of $\mu$ between 0.220 and 0.306 in
increments of 0.001.  Concerning the initial conditions in $\rho_0$, denoting
the relative starting position of the planet, we use values between 0.272 and
0.502 in increments of 0.001.  We show the
$1 \sigma$ region in $\rho_0$ between 0.344 and 0.418 where $\nu$ Oct is proposed
to exist by a grey shaded region; see Sect.~\ref{sec:SPS}.  Between the
two bounding values regarding $\rho_0$, there
exists a region of stability as well as a marginal stability limit.  The limit
of marginal stability is consistent with previous findings by \cite{hol99} for
the case with the planetary eccentricity taken into account.  This region of
marginal stability is more pronounced in the 3 o'clock starting configuration;
it begins at $\rho_0\approx 0.315$.  The marginal stability region decreases
as $\mu$ increases, which is expected because the perturbing mass is becoming
more significant as the mass ratio $\mu$ increases.  There is a noticeable
favoritism towards the 3 o'clock starting position as indicated by the survival
counts given in Table~\ref{tab:SC}.  This outcome is expected as the perturbing
mass exerts a lesser force on the planet if initially placed at the 3 o'clock
position.

Figure~\ref{fig:Fig03} shows the result for the retrograde configuration
for 1,000 binary orbits considering both the 3 o'clock and 9 o'clock planetary
starting positions.  Due to the larger number of conditions that survived during 
this time interval we have reduced the bounds of initial conditions following
the bounding values of \cite{ebe10b}.  Here the bounding values for $\mu$
are given as 0.2593 and 0.2908 with increments of 0.0001.  The increment
in $\mu$ was reduced to allow better insight into the role of chaos in
this type of configuration.  The initial conditions of $\rho_0$ begin with
0.353 and end with 0.416 in increments of 0.001, which allows to cover the
most pronounced features of the retrograde configuration.

Figure~\ref{fig:Fig03} demonstrates a broad region of stability within the
selected parameter space.  The 9 o'clock starting position has many regions
of bounded chaos.  These regions of bounded chaos exhibit locally less
negative values for MLE, which however do not exceed the previously determined
cutoff value of log~$\lambda_{\rm max}=-0.82$; see Sect.~\ref{sec:Criteria}.  The
nature of these orbits and the connection to the MLE will be discussed in
Sect.~\ref{sec:CStud}.  The greatest region of bounded chaos is obtained for
$\rho_0\approx0.41$.  This region decreases in $\rho_0$ as $\mu$ increases
for the same reason as the change in the marginal stability limit in
Fig.~\ref{fig:Fig02}.  The 3 o'clock starting position shows similar
features but there exists a region of instability with this starting position
at $\rho_0\approx0.41$.  This region coincides with the extended region of
bounded chaos in the 9 o'clock starting position.  The nature of this region
can be deduced from the parameters given by \cite{ram09}; see
Table~\ref{tab:Parameters}.  There exists a 5:2 resonance of the planet with
the binary that places the suggested planet moving towards the barycentre
at the periastron point of the binary inducing an instability.  For the
9 o'clock position the suggested planet would be moving away which would
provide a large perturbation but insufficient for producing an instability.

\begin{figure*}

\centering
\begin{tabular}{c}
\includegraphics[trim = 0mm 1mm 1mm 1mm, clip, width=0.750\linewidth]{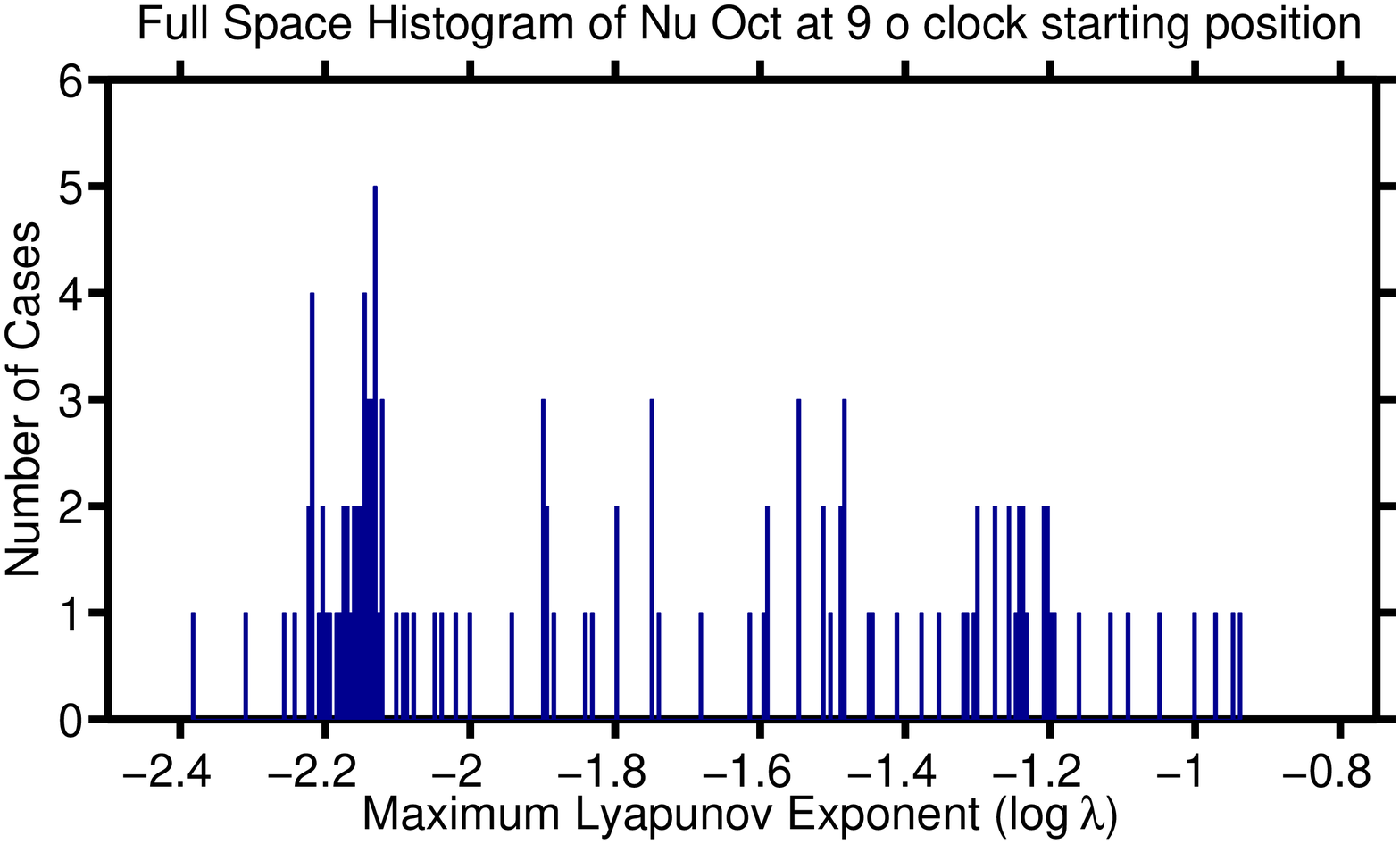} \\
\includegraphics[trim = 0mm 1mm 1mm 1mm, clip, width=0.825\linewidth]{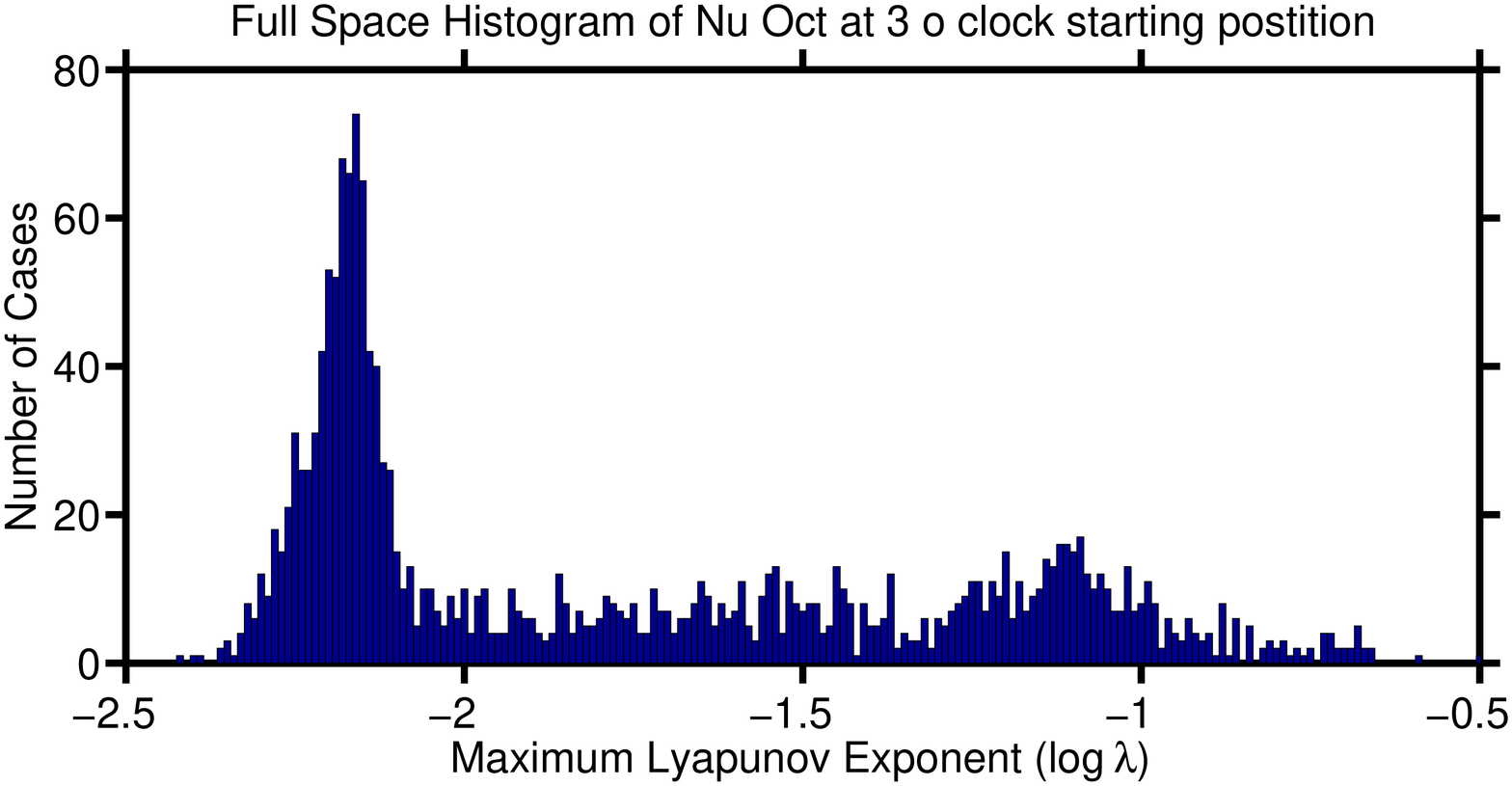}
\end{tabular}
\caption{Histograms of configurations that survived 1000 binary orbits in prograde motion.
Note the significant difference in scale regarding both the $x$ and $y$-axis.}
\label{fig:Fig04}
\end{figure*}

Figure~\ref{fig:Fig04} provides the histograms of simulations pertaining to both
starting positions for the prograde configuration.  These histograms indicate the
number of surviving cases related to the array of initial conditions for the
parameter space detailed in Fig.~\ref{fig:Fig02}.  The bin width is 0.01 in the
logarithmic MLE scale, corresponding to $1.47\times10^{-4}$ in the linear MLE scale.
Although there is not a case that survives 10,000 binary orbits in the 1 $\sigma$ region, we provide these
histograms to demonstrate what different initial conditions entail near the
lower and upper uncertainty limits.  Figure~\ref{fig:Fig02} shows that all the
surviving cases cluster towards the lower bound in $\rho_0$.  Furthermore, it is
obtained that the 9 o'clock starting position has far fewer surviving initial
conditions than the 3 o'clock starting position.  This is also reflected in the
histograms.  The histogram of the 9 o'clock position is sparse; without the aid
of the 3 o'clock histogram there would be too few counts to provide substantial
information.  But inspecting the 3 o'clock histogram reveals that there is a
Gaussian-type distribution about the logarithmic MLE value of $-2.2$.  This
indicates that for the limited simulation time adopted for this part of our study,
there is a very small number of cases indicating stability for the prograde
configuration.  However, these cases are near the $2 \sigma$ or $3 \sigma$ limit
concerning the permitted values of the planetary position $\rho_0$ as discussed
in Sect.~\ref{sec:StatS}.

\subsection{Case Studies}
\label{sec:CStud}

\begin{figure*}

\centering
\begin{tabular}{ccc}
\epsfig{file=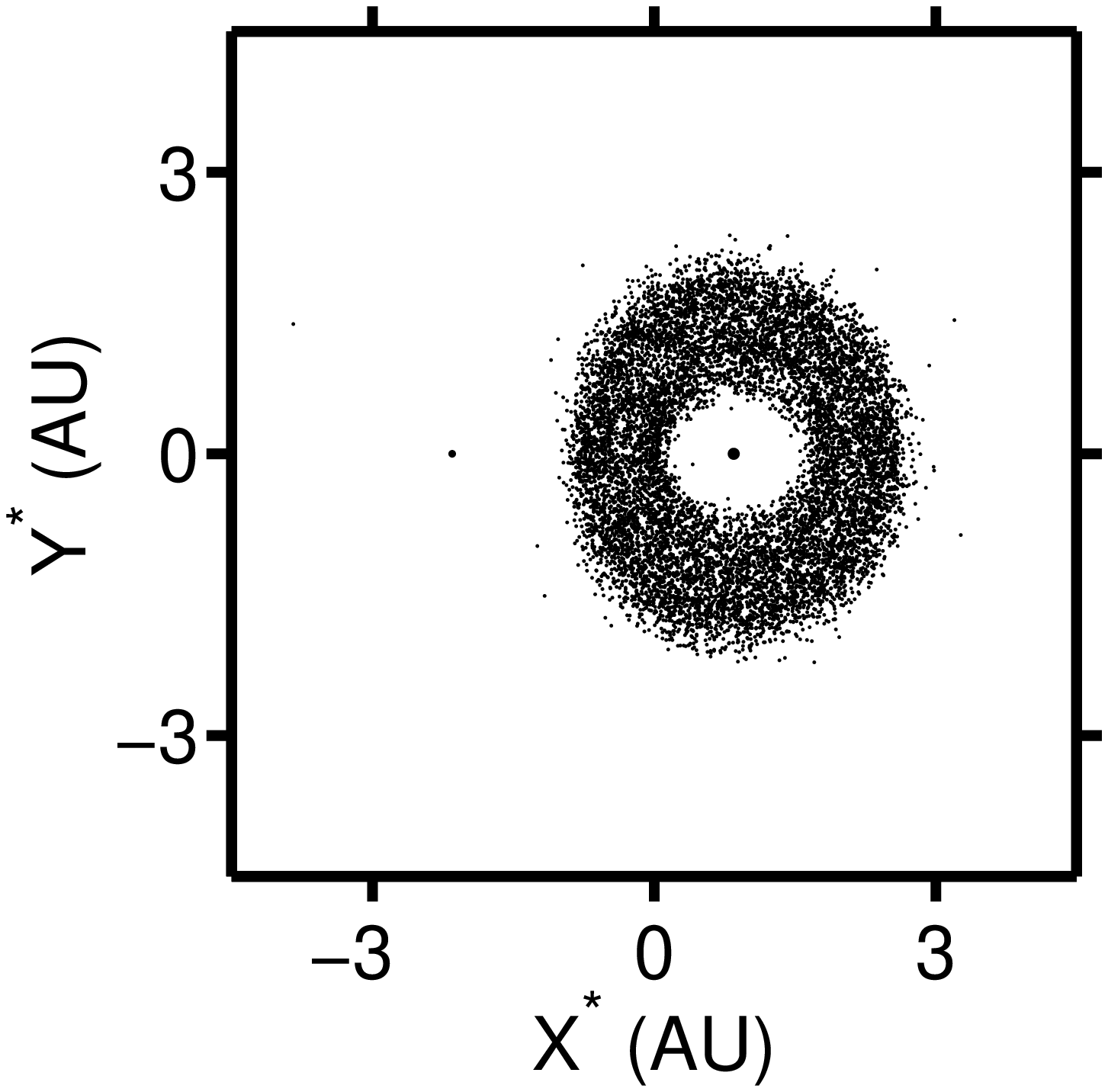,width=0.3\linewidth,height=0.3\linewidth}&
\epsfig{file=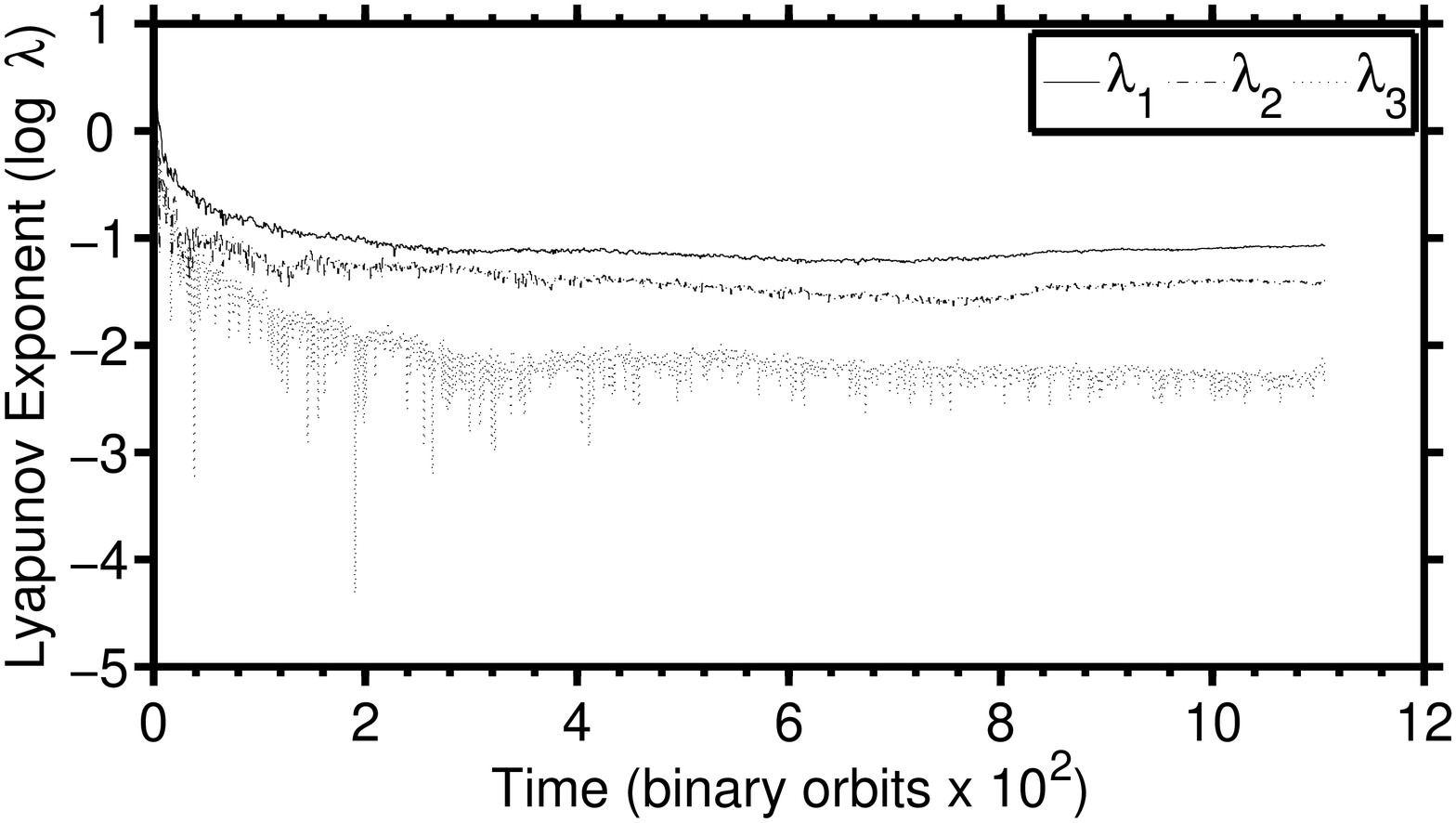,width=0.35\linewidth,height=0.3\linewidth}&
\epsfig{file=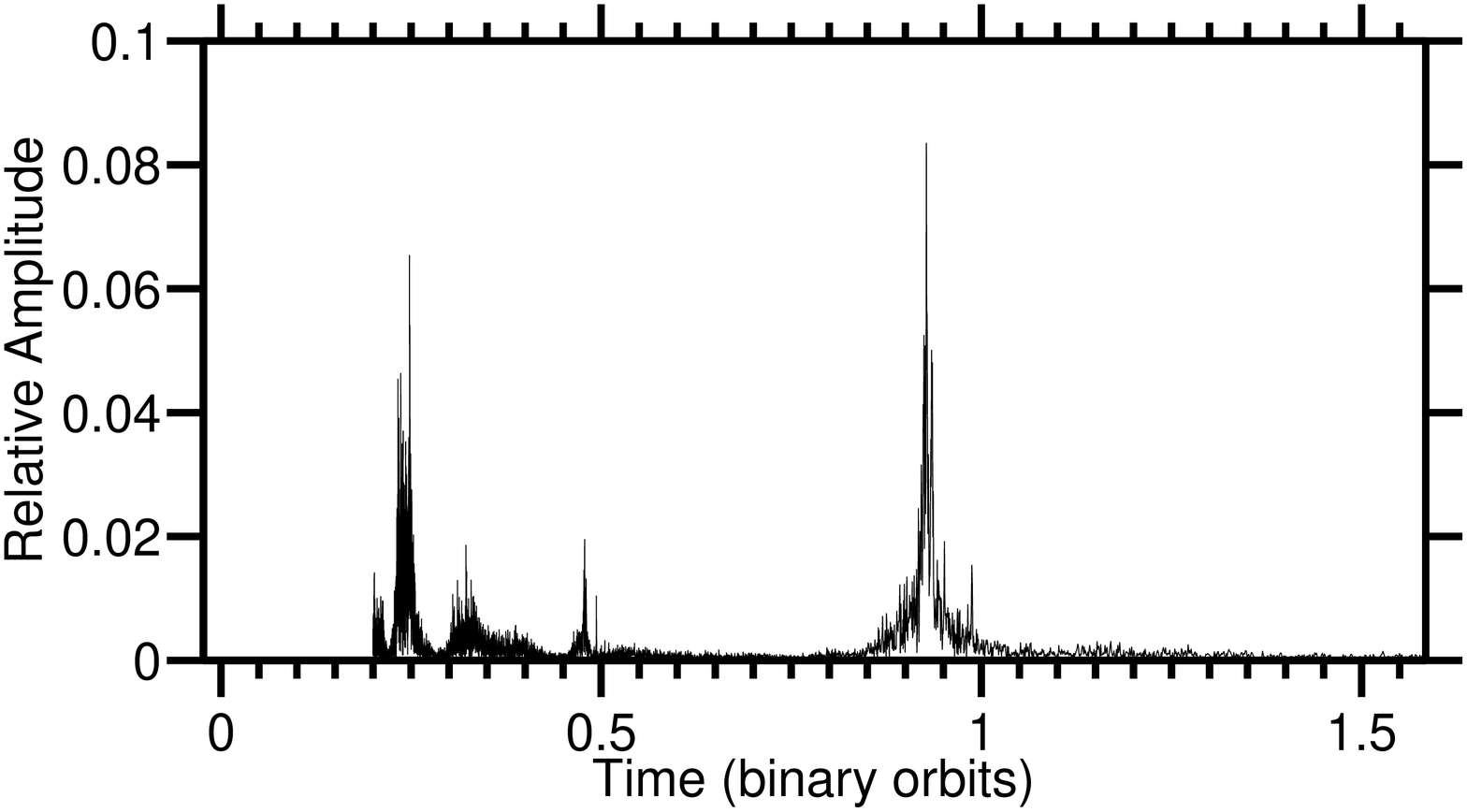,width=0.35\linewidth,height=0.3\linewidth}\\
\end{tabular}
\caption{Case study of planetary motion with the planet placed in the
3 o'clock position and in retrograde motion.
This case displays the conditions for $\mu = 0.2825$ and $\rho_0 = 0.400$.}
\label{fig:Fig05}
\end{figure*}

\begin{figure*}

\centering
\begin{tabular}{ccc}
\epsfig{file=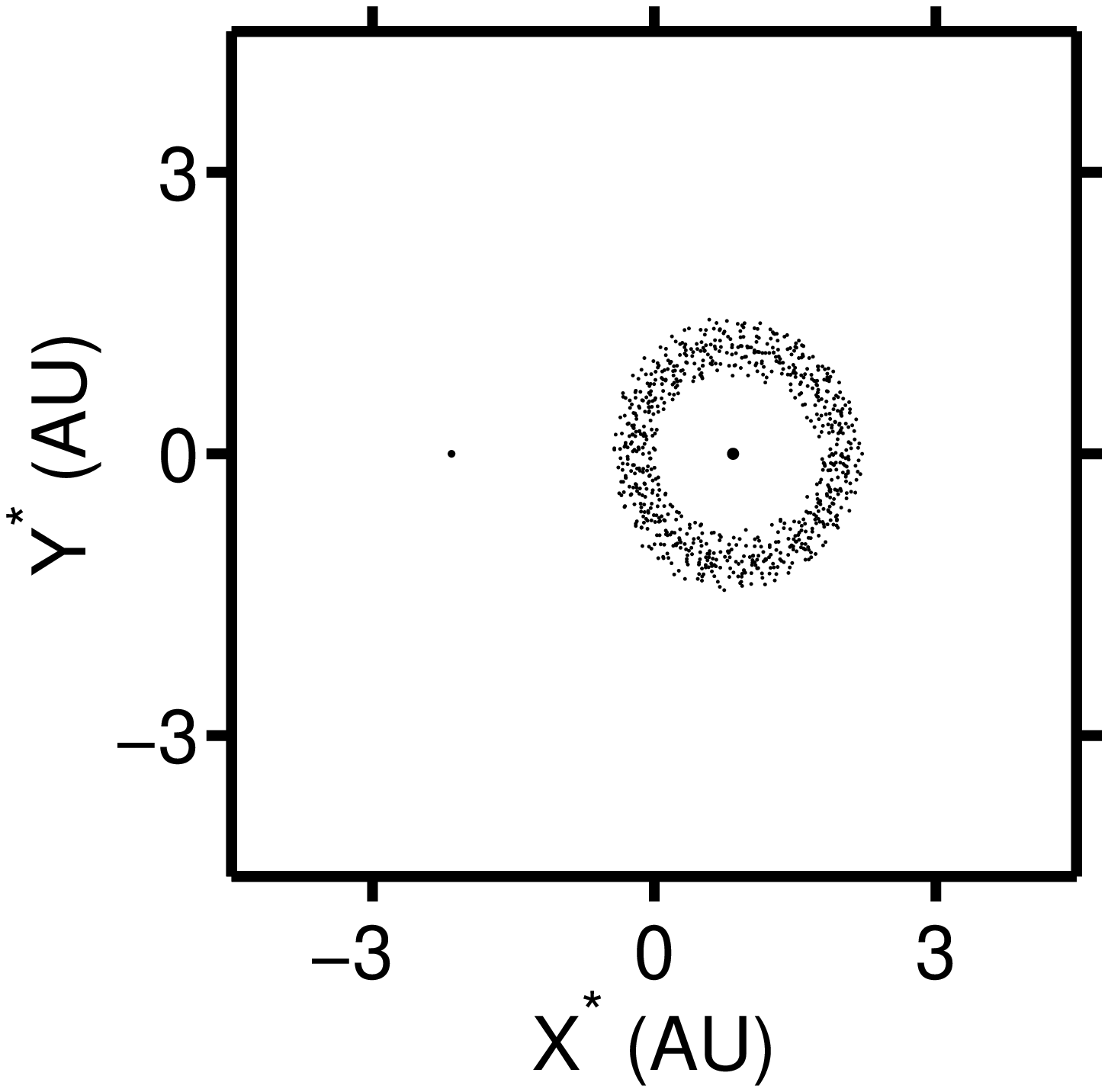,width=0.3\linewidth,height=0.3\linewidth}&
\epsfig{file=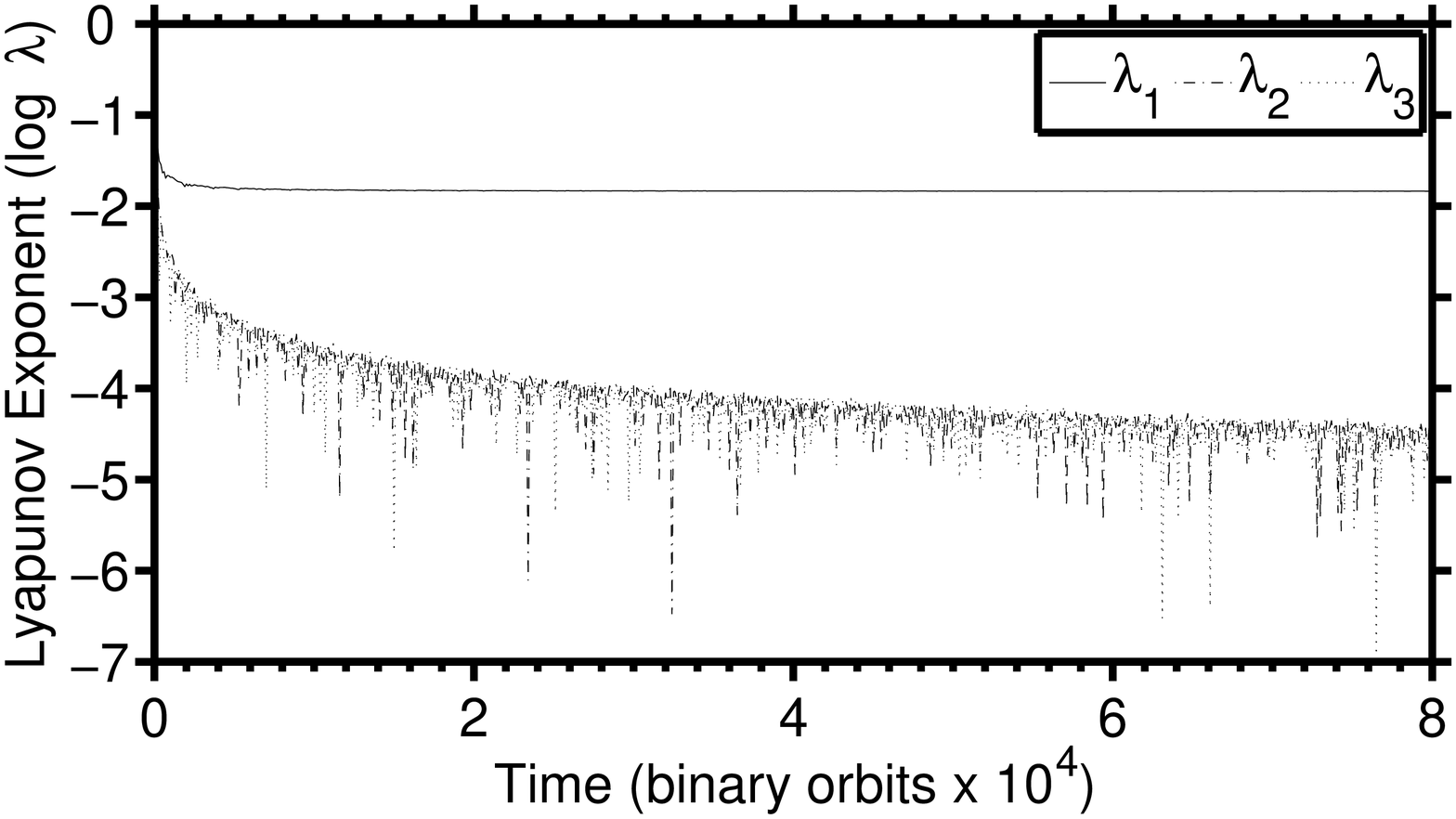,width=0.35\linewidth,height=0.3\linewidth}&
\epsfig{file=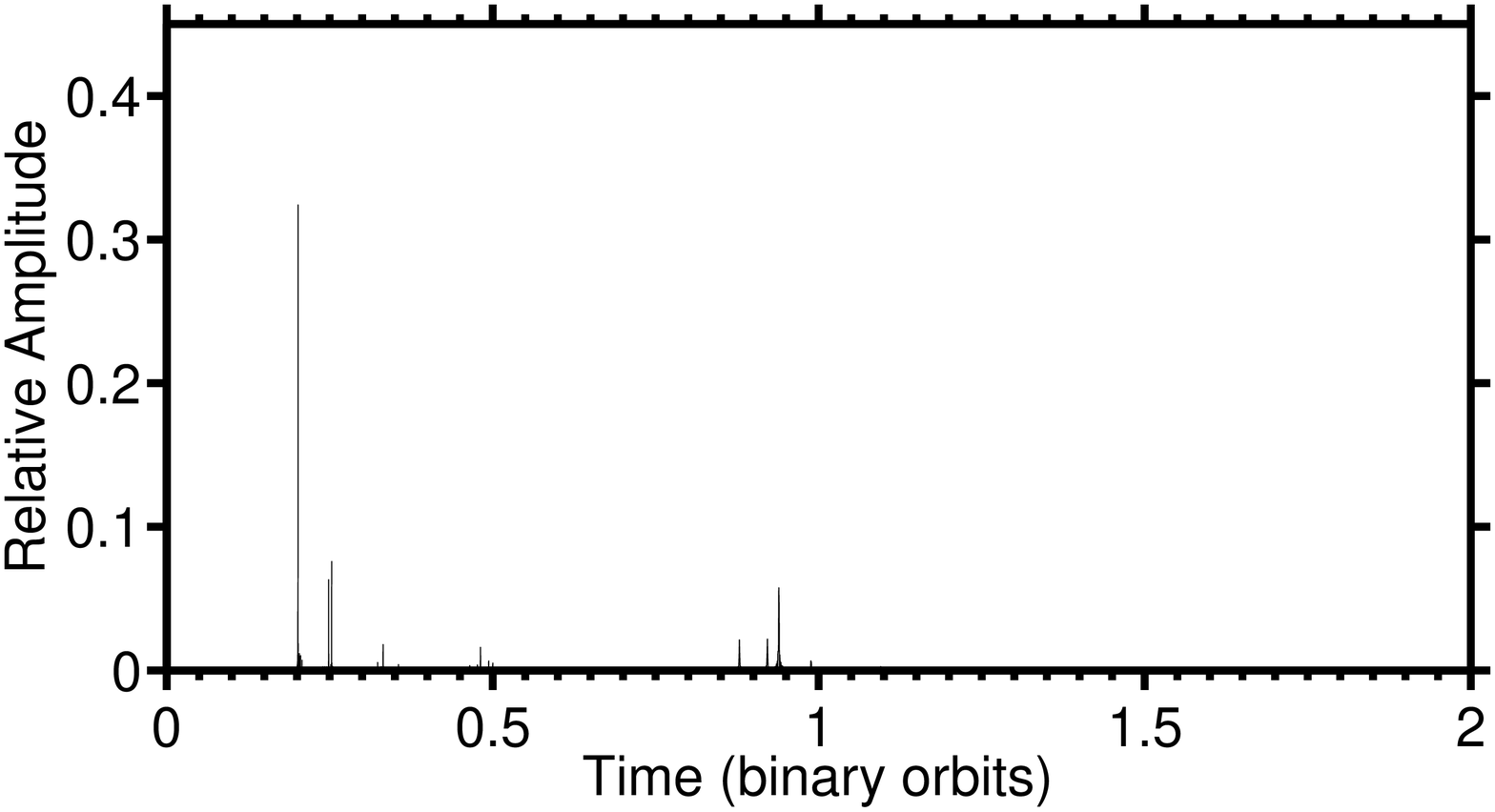,width=0.35\linewidth,height=0.3\linewidth}\\
\end{tabular}
\caption{Case study of planetary motion with the planet placed in the
3 o'clock position and in retrograde motion.
This case displays the conditions for $\mu = 0.2805$ and $\rho_0 = 0.358$.}
\label{fig:Fig06}
\end{figure*}

\begin{figure*}

\centering
\begin{tabular}{ccc}
\epsfig{file=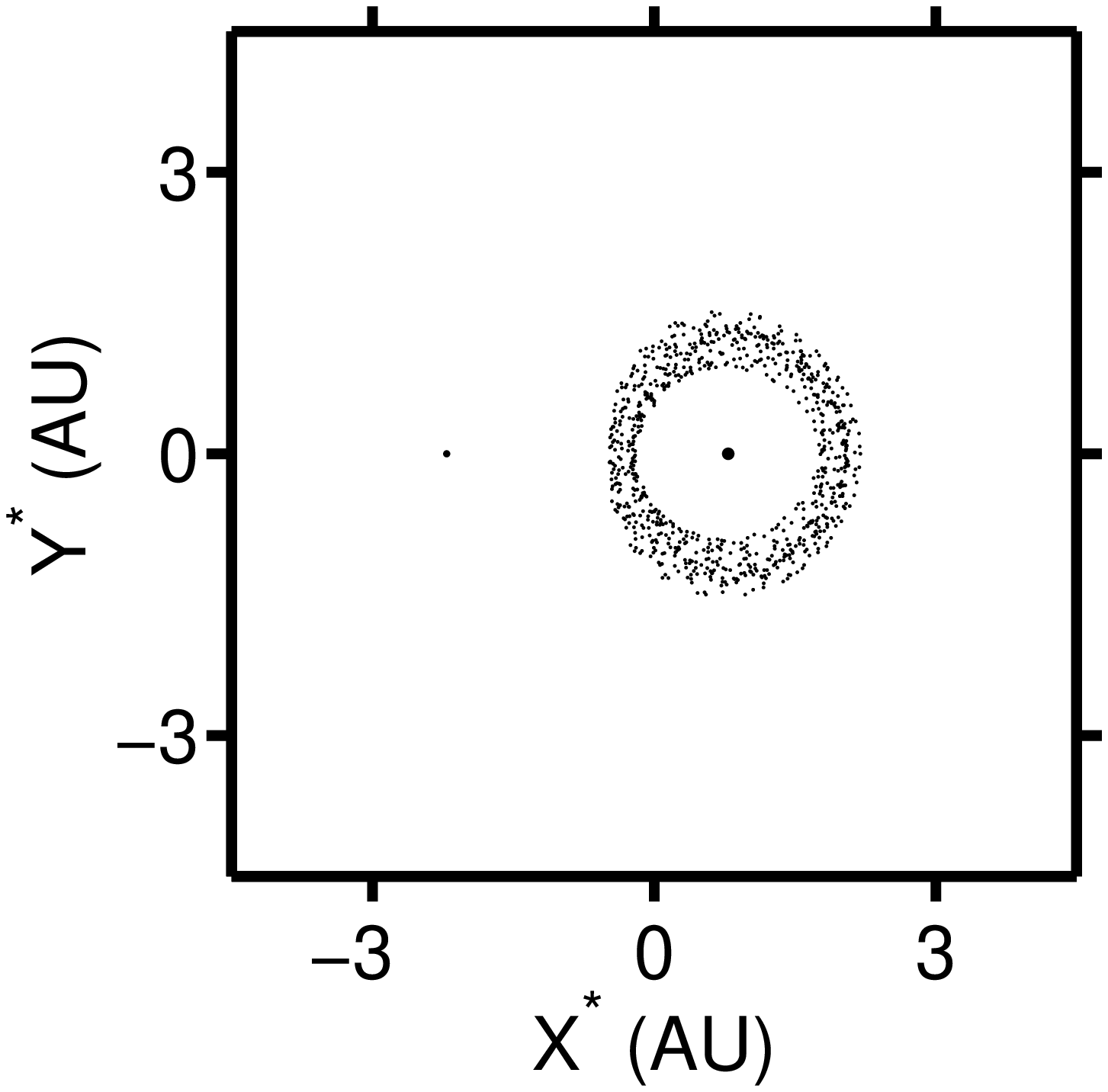,width=0.3\linewidth,height=0.3\linewidth}&
\epsfig{file=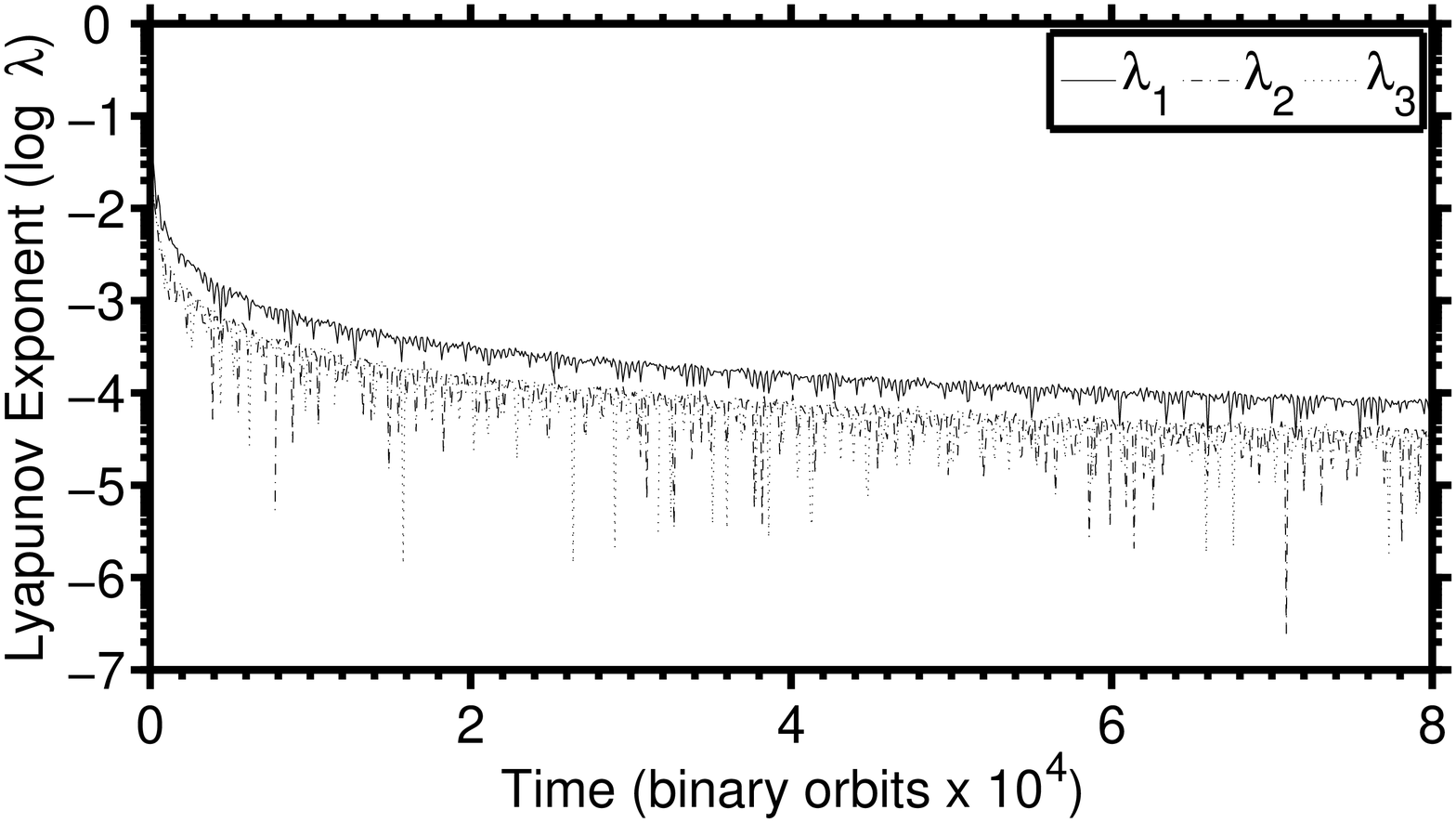,width=0.35\linewidth,height=0.3\linewidth}&
\epsfig{file=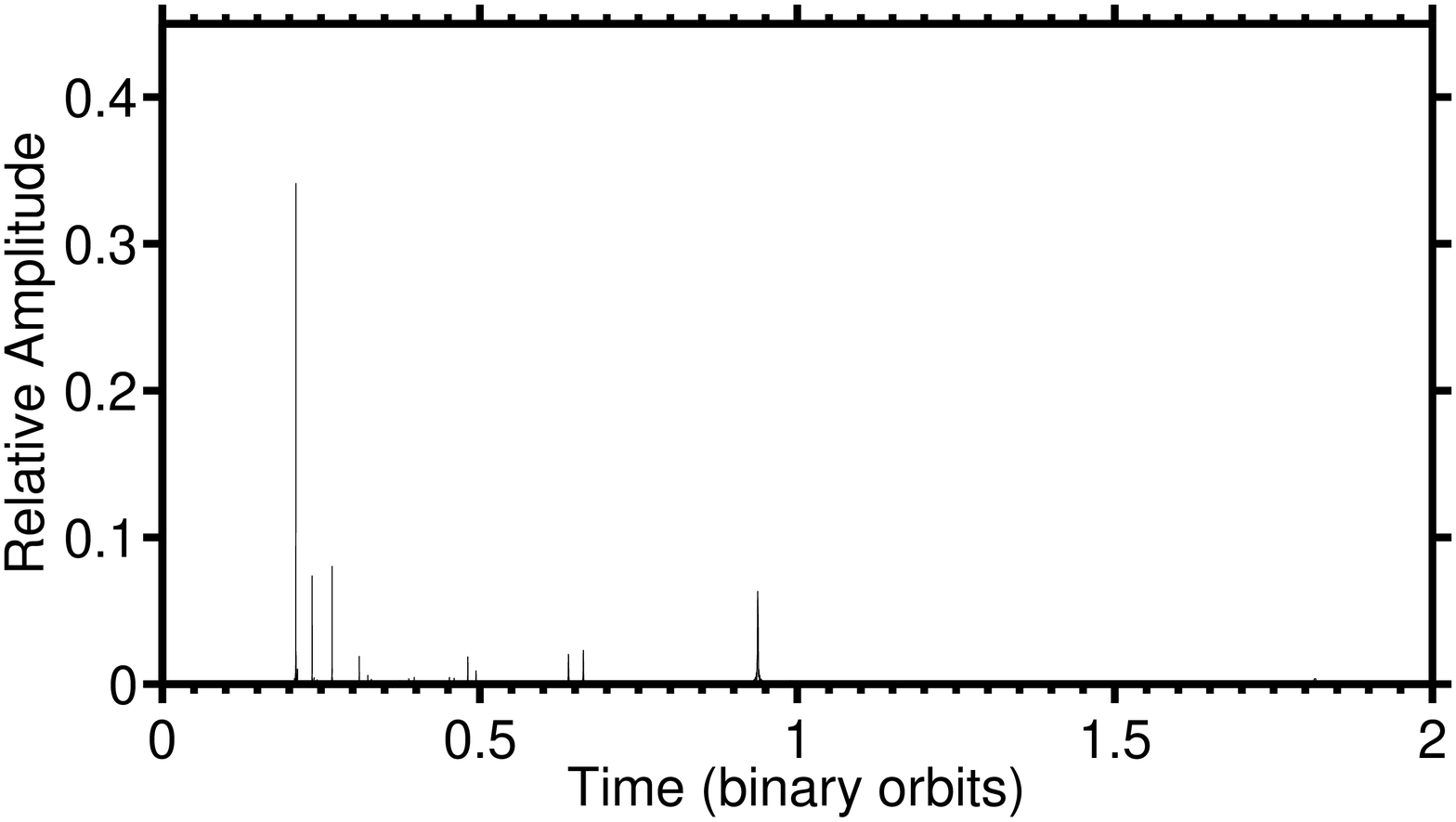,width=0.35\linewidth,height=0.3\linewidth}\\
\end{tabular}
\caption{Case study of planetary motion with the planet placed in the
3 o'clock position and in retrograde motion.
This case displays the conditions for $\mu = 0.2630$ and $\rho_0 = 0.374$.}
\label{fig:Fig07}
\end{figure*}

\begin{figure*}

\centering
\begin{tabular}{ccc}
\epsfig{file=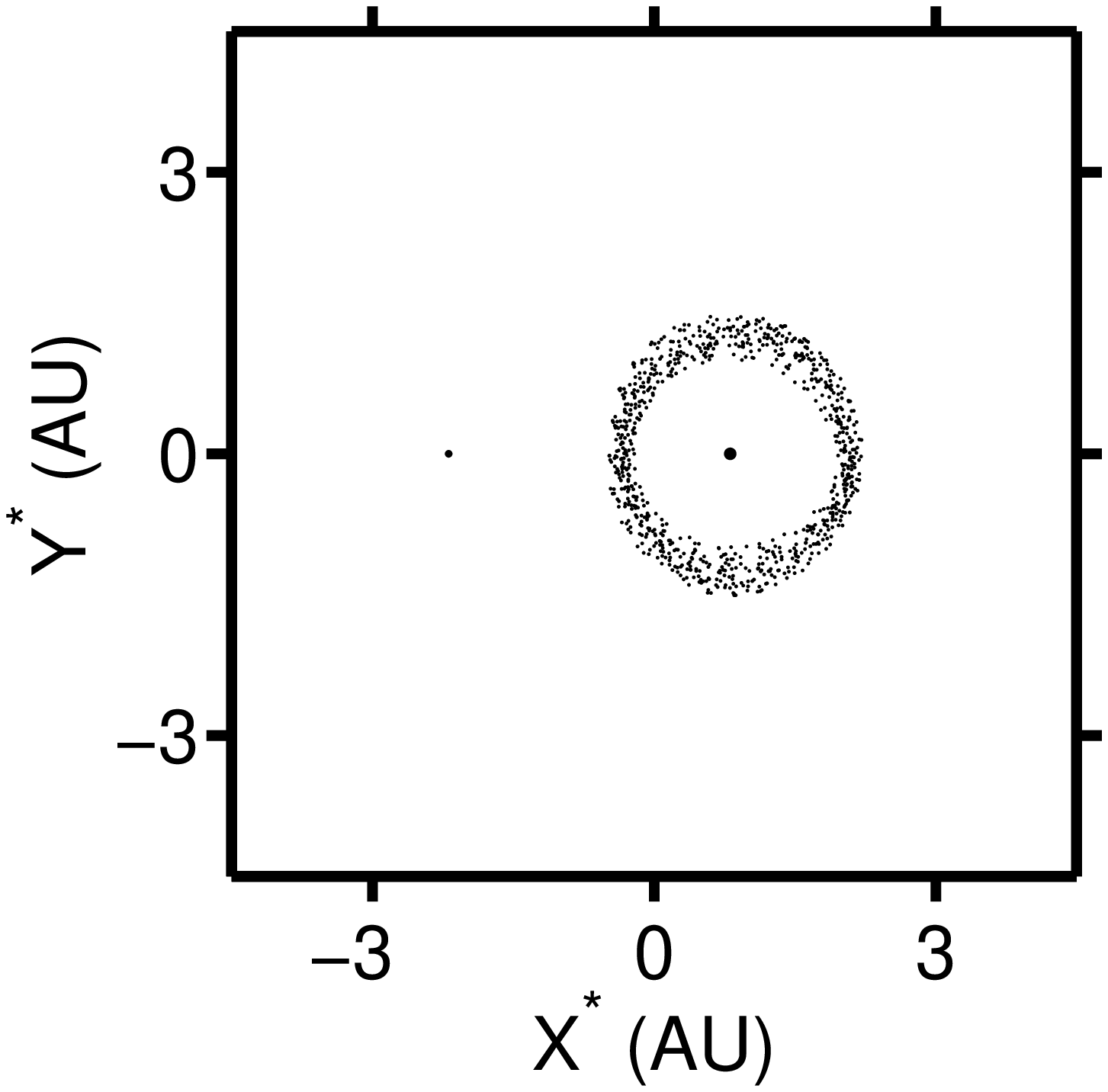,width=0.3\linewidth,height=0.3\linewidth}&
\epsfig{file=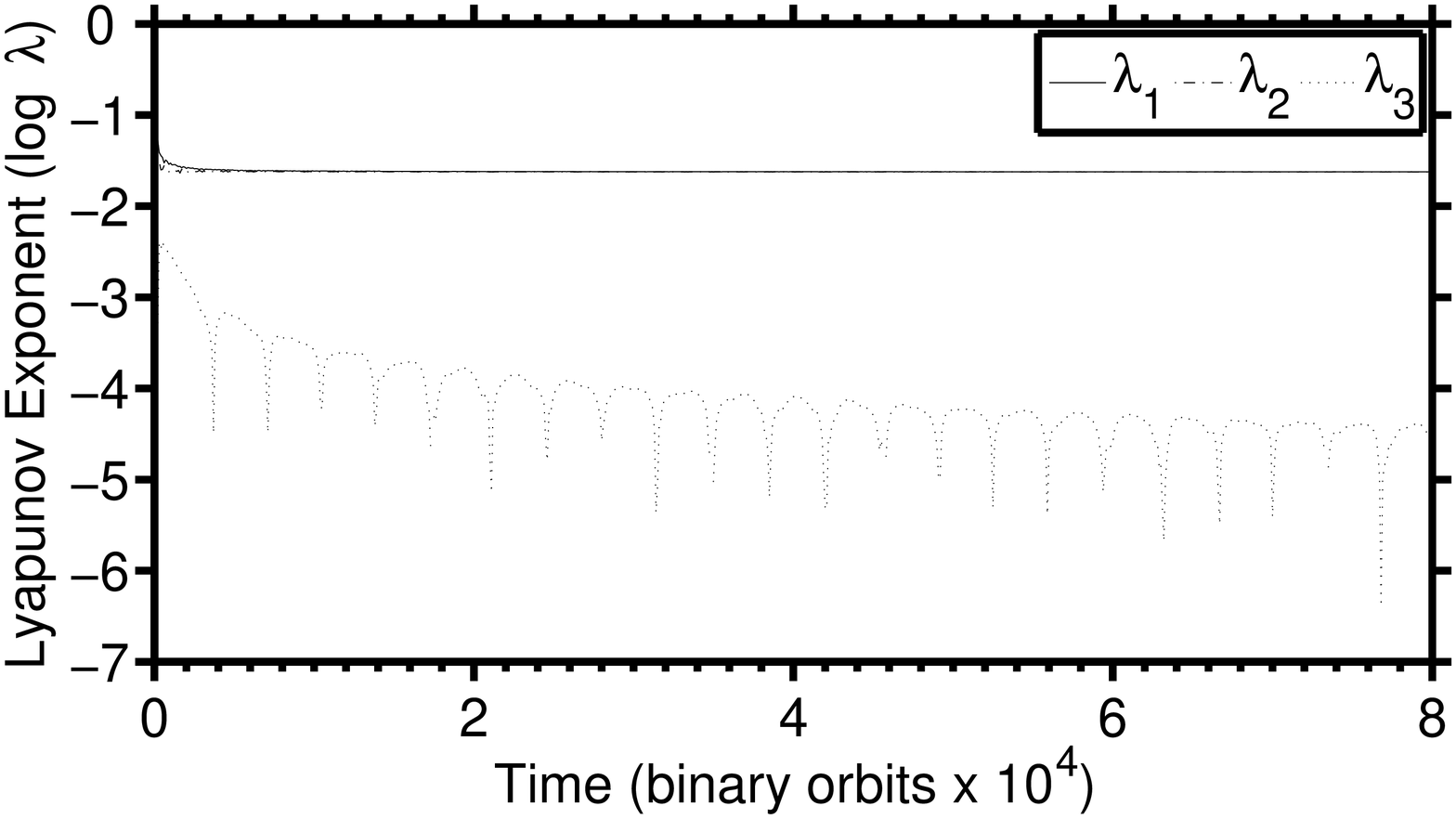,width=0.35\linewidth,height=0.3\linewidth}&
\epsfig{file=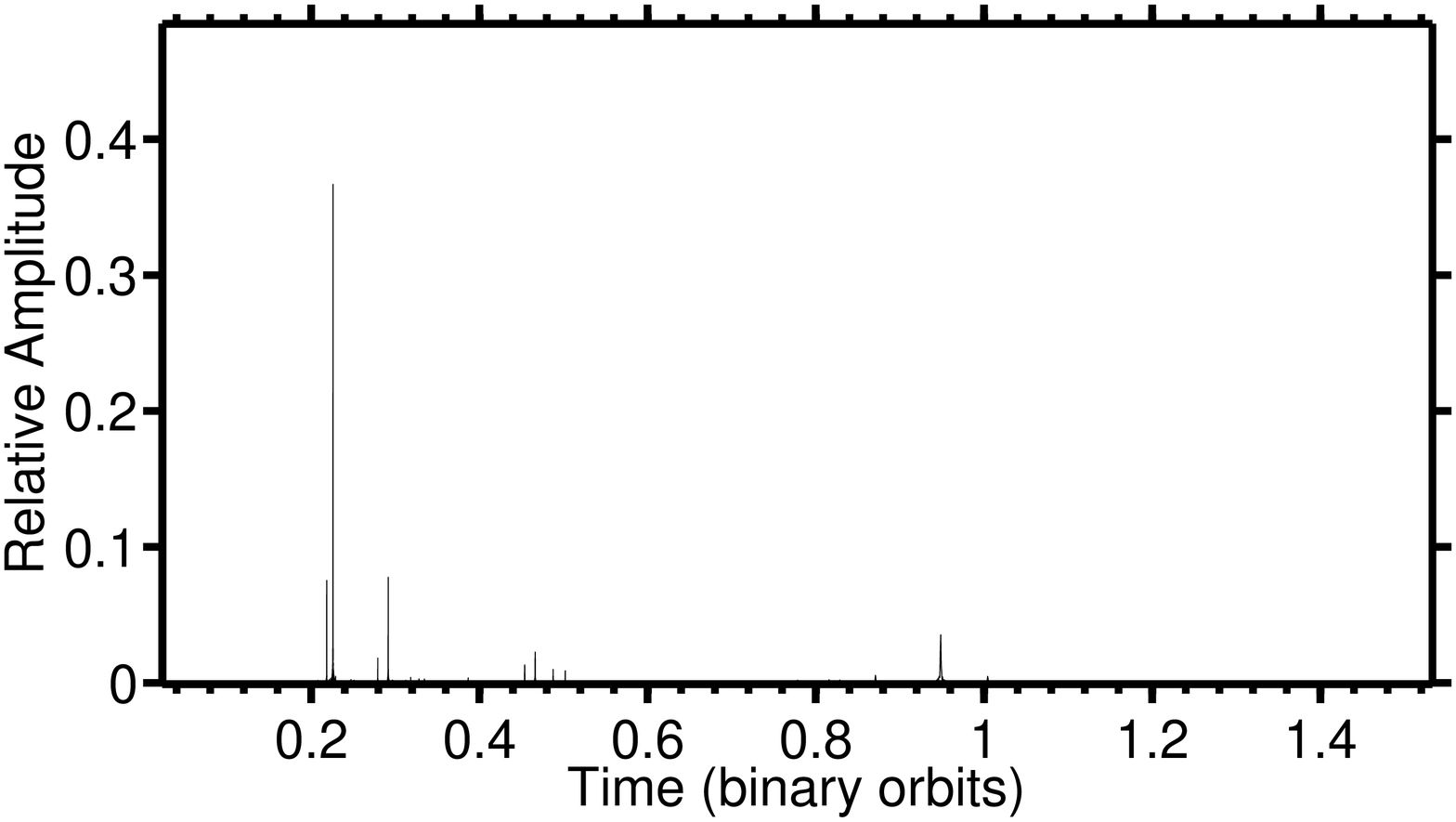,width=0.35\linewidth,height=0.3\linewidth}\\
\end{tabular}
\caption{Case study of planetary motion with the planet placed in the
9 o'clock position and in retrograde motion.
This case displays the conditions for $\mu = 0.2696$ and $\rho_0 = 0.401$.}
\label{fig:Fig08}
\end{figure*}

\begin{figure*}

\centering
\begin{tabular}{ccc}
\epsfig{file=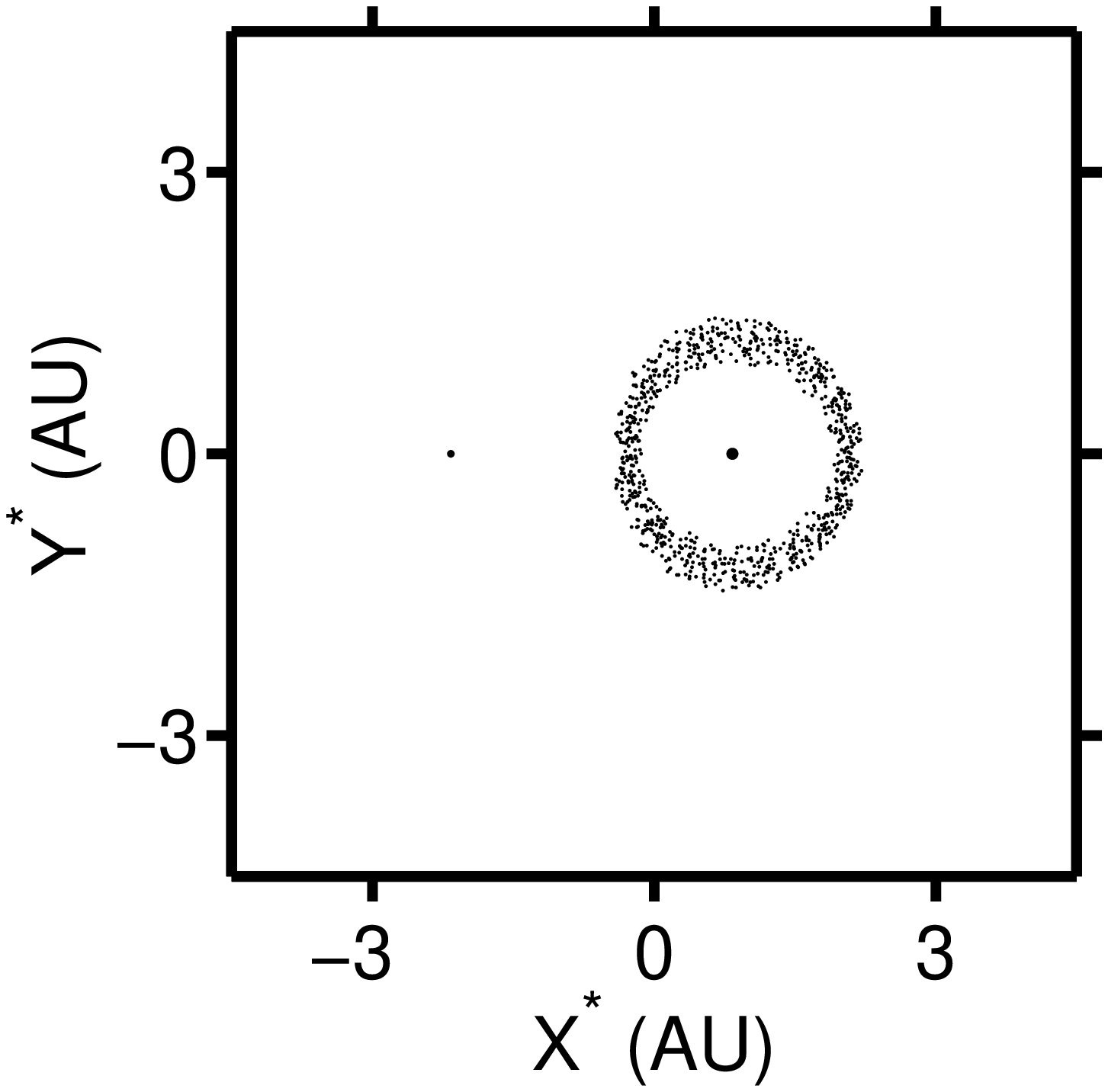,width=0.3\linewidth,height=0.3\linewidth}&
\epsfig{file=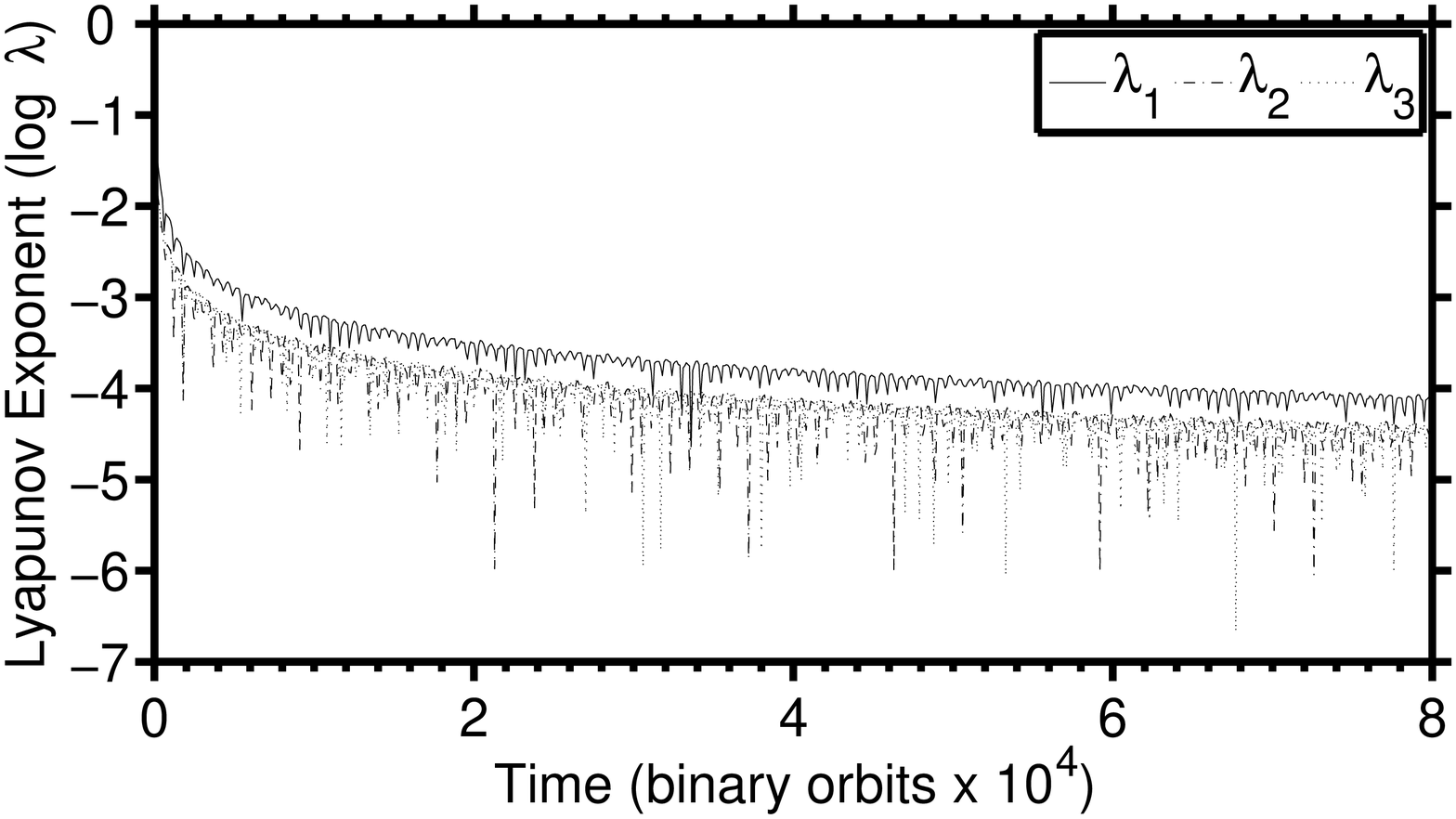,width=0.35\linewidth,height=0.3\linewidth}&
\epsfig{file=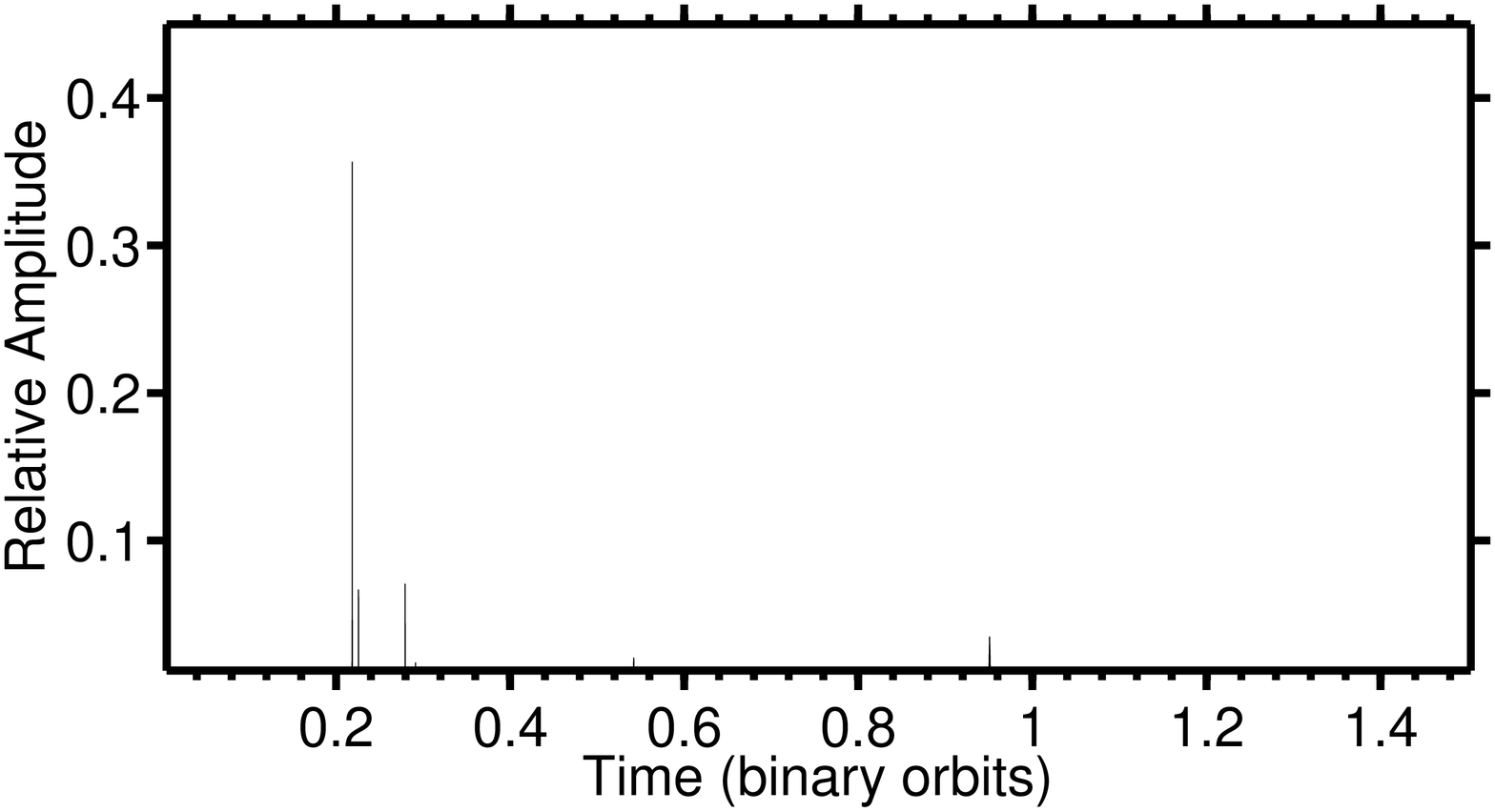,width=0.35\linewidth,height=0.3\linewidth}\\
\end{tabular}
\caption{Case study of planetary motion with the planet placed in the
9 o'clock position and in retrograde motion.
This case displays the conditions for $\mu = 0.2780$ and $\rho_0 = 0.388$.}
\label{fig:Fig09}
\end{figure*}

In the following we probe some orbits of retrograde configurations in more detail;
see Figs.~\ref{fig:Fig05} to \ref{fig:Fig09} for information.
These figures show from left to right the respective orbital diagram in a rotating
reference frame, the associated Lyapunov spectrum, and the Fourier periodogram.
Through these diagrams we can assess the nature of these orbits to
determine stability and measure the chaos within the system.

Figure~\ref{fig:Fig05} with $(\mu,\rho_0) = (0.2825,0.400)$ is provided to
show a case of instability.  In this case we find that the orbit presents
itself as a wide annulus and already at first glance appears chaotic.  The Lyapunov
spectrum reveals that the MLE levels off at log~$\lambda = -1$ fairly quickly, which is near
the proposed cutoff value for unstable chaotic behaviour.  For the timespan between 300
and 800 binary orbits, the MLE fluctuates until it finally pushes upward
to $\lambda_{\rm max}$.  The Fourier periodogram shows that this case is truly
chaotic due to the amount of noise presented by the periodogram.  There are
many periods which may promote the outcome of instability via resonance overlap
\citep[e.g.,][]{mud06,mar07}.  The proposed planet is approaching the barycentre
when the binary is at its periastron position, which should be considered
the main cause of this instability.

Figures~\ref{fig:Fig06} and \ref{fig:Fig07} with $(\mu,\rho_0) = (0.2805,0.358)$
and $(0.2630,0.374)$, respectively, exhibit similar features between each other in the orbit
diagram and periodogram.  However, there is a difference in the respective Lyapunov
spectra.  Both cases show trends of stability, although Fig.~\ref{fig:Fig06}
demonstrates bounded chaos.  It presents a different nature than the previous case
of instability (see Fig.~\ref{fig:Fig05}) as the MLE settles near log~$\lambda = -1.8$
and does not fluctuate enough to reach $\lambda_{\rm max}$.  Both cases given
as Figs.~\ref{fig:Fig06} and \ref{fig:Fig07} represent orbits that form annuli;
however, the annulus widths are less than that of the unstable case depicted in
Fig.~\ref{fig:Fig05}.  They are thus classified as stable with Fig.~\ref{fig:Fig06}
being chaotic and Fig.~\ref{fig:Fig07} being non-chaotic.

Figures~\ref{fig:Fig08} and \ref{fig:Fig09} with $(\mu,\rho_0) = (0.2696,0.401)$
and $(0.2780,0.388)$, respectively, reveal similar features compared to the
previous set of figures regarding all three diagrams.  These cases are provided
to demonstrate the similarities between the 3 o'clock starting positions depicted
in Figs.~\ref{fig:Fig06} and \ref{fig:Fig07} and the 9 o'clock starting positions
depicted in Figs.~\ref{fig:Fig08} and \ref{fig:Fig09}.  The comparison between
Figs.~\ref{fig:Fig06} and \ref{fig:Fig08} indicates a similar orbit diagram and
periodogram, but the limiting values of the MLE settle at a slightly more negative value
in Fig.~\ref{fig:Fig06}.  Figures~\ref{fig:Fig07} and \ref{fig:Fig09} supply
similar orbital diagrams but there are minor differences in the periodograms and
Lypaunov spectra.  These figures display the same behaviour in MLE but there
is a difference in the periodic nature of the decrease in MLE, which is also
revealed by the difference in the number of peaks in the corresponding
periodograms.

\begin{table}

\begin{center}
\caption{Statistical Probability of Prograde Planetary Orbits$^a$.}
\vspace{0.05in}
\vspace{0.05in}
\begin{tabular}{l c c c c}
\hline
\hline
\noalign{\vspace{0.03in}}
Mass Ratio &  \multicolumn{2}{c}{9 o'clock Position} &  \multicolumn{2}{c}{3 o'clock Position}   \\
\noalign{\vspace{0.03in}}
\hline
\noalign{\vspace{0.03in}}
$\mu$   &  $\sigma$  &  Probability (\%)  &  $\sigma$  & Probability (\%)  \\
\noalign{\vspace{0.03in}}
\hline
\noalign{\vspace{0.03in}}
0.306   &   $\ldots$ &  $\ldots$  &  2.30  &  1.1  \\
0.263   &   $\ldots$ &  $\ldots$  &  2.11  &  1.7  \\
0.220   &   2.95     &  0.16      &  1.89  &  2.9  \\
\noalign{\vspace{0.05in}} \hline
\end{tabular}
\label{tab:StatP}
\vspace{0.05in}
\begin{enumerate}
\item[$^a$] Based on 10,000 binary orbits; see Fig.~\ref{fig:Fig02}
\end{enumerate}
\end{center}

\end{table}

\subsection{Statistical Analysis of Orbital Stability}
\label{sec:StatS}

A significant component of the present study is to provide a
statistical analysis regarding whether or not the suggested planet
in the $\nu$~Octantis system is able to exist in a prograde or
retrograde orbit.  Previously, we reported on the general behaviour of
orbital stability for the set of simulations with focus on the MLE
(see Sect.~\ref{sec:Criteria}).
In Figs.~\ref{fig:Fig02} and~\ref{fig:Fig03}, the various values
of the MLE were given as function of the mass ratio $\mu$ and the
initial planetary distance parameter $\rho_0$.  The bounding values
for $\mu$ were given as $\left[0.220, 0.306\right]$,
i.e., $0.263 \pm 0.043$, and for $\rho_0$ as $\left[0.344, 0.418\right]$,
i.e., $0.381 \pm 0.037$.  We found that based on the MLE as well as
a number count of surviving cases (see Sect.~\ref{sec:ModelS}) that the
existence of the suggested planet in a retrograde orbit is almost
certainly possible, whereas in a prograde orbit it is virtually
impossible.

For prograde orbits based on the ``landscape of survival" (see
Fig.~\ref{fig:Fig02}), another type of analysis can be given, see
Table~\ref{tab:StatP}. 
We conclude that for the 9 o'clock planetary starting position
and for a relatively high stellar mass ratios, i.e., $\mu \gta 0.25$,
the planet must start very close to the star, i.e., that the
associated probability for $\rho_0$ is way outside the statistical
$3 \sigma$ limit.  Only for relatively small stellar mass ratios,
i.e., $\mu \simeq 0.22$, the planetary starting distance $\rho_0$
would be barely consistent with the $3 \sigma$ limit, amounting
to a statistical probability of 0.16\%.  For a 3 o'clock planetary
starting position, these probabilities pertaining to the permitted
values of $\rho_0$ are moderately increased.  It is found that the
associated probabilities for $\mu$ given as 0.306, 0.263, and 0.220
are 1.1\%, 1.7\%, and 2.9\%, respectively.

Moreover, it should be noted that this statistical analysis refers
to a ``landscape of survival" that has been obtained for a timespan
of 10,000 binary orbits, corresponding to 29,000 years.  An analysis
of the evolution of that landscape shows that it tends to dissipate
if larger timespans are adopted.  Even for timespans of 10,000 binary orbits, there
are many parameter combinations $(\mu,\rho_0)$ where, concerning
prograde orbits, no orbital stability is found even if $\rho$ values
are chosen at or beyond the statistical $3 \sigma$ level.

\begin{figure*}

\centering
\begin{tabular}{cc}
\includegraphics[trim = 0mm 1mm 1mm 1mm, clip, width=0.85\linewidth]{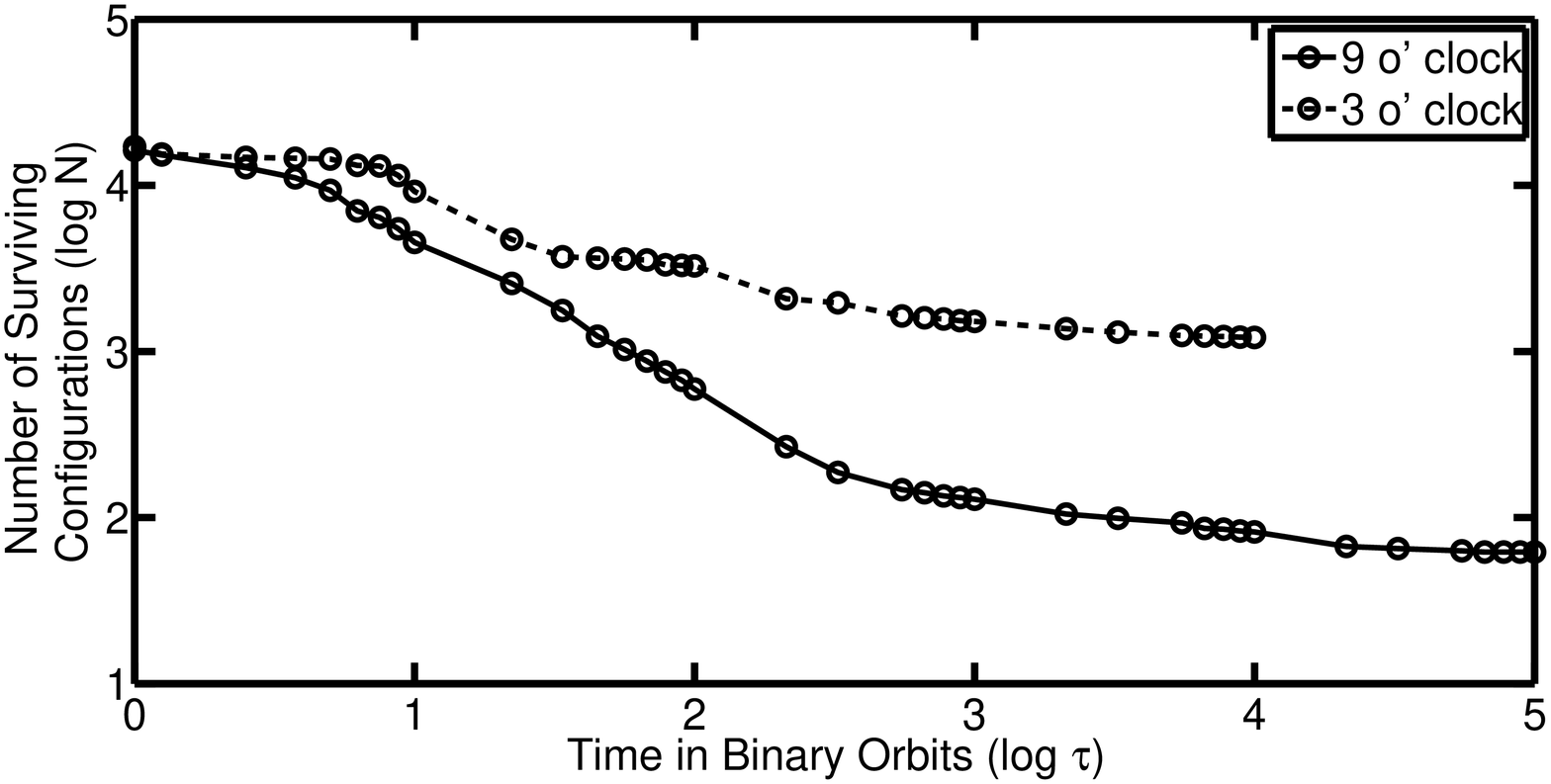}
\end{tabular}
\caption{Logarithmic representation of the surviving number of configurations
(prograde orbits) as function of time given in units of completed binary orbits.}
\label{fig:Fig10}
\end{figure*}

Another perspective is offered through Fig.~\ref{fig:Fig10}, which
provides additional insight into the statistical nature of our simulations.
It addresses the view of long term stability.
Figure~\ref{fig:Fig10} depicts two curves representing the number of surviving
configurations as a function of time for each starting position in the prograde
configuration.  The features of this logarithmic plot show us that for the first
10 binary orbits there is not a preference towards stability of either starting
position.  After 10 binary orbits the 9 o'clock position shows a significantly
steeper trend of instability as indicated by the slope of the respective curve.
The 3 o'clock position also shows a trend toward instability but it is
asymptotically approaching a limit of 10$^3$ much more smoothly.  With these
limits, we can estimate the likelihood for a prograde planetary configuration
to be extremely small in the framework of long-term simulations.  However, for
a very large range of $(\mu,\rho_0)$ combinations, the stability of retrograde
planetary configuration appears almost certainly guaranteed, even for timescales
beyond $10^5$ binary orbits.

\section{Conclusions}

This study offers detailed insights into the principal possibility of
prograde and retrograde orbits for the suggested planet in the
$\nu$~Octantis system, if confirmed through future observations.
A previous study by \cite{ebe10b} concluded that the planet is most
likely stable if assumed to be in a retrograde orbit with respect to
the secondary system component.  In the present work, we were able to
confirm this theoretical finding, while taking into account the
observationally deduced uncertainty ranges of the orbital parameters
for the stellar components and the suggested planet as well as
different mass ratios of the stellar components.

In addition, we also employed additional mathematical methods,
particularly the behaviour of the maximum Lyapunov exponent (MLE).
We found that virtually all cases of prograde orbit became unstable
over a short amount of time; the best cases for survival (although
barely significant in a statistical framework) were found if a
relatively small mass ratio for the two stellar components was assumed
and if the initial configuration was set for the planet to start at
the 3 o'clock position.  This later preference is also consistent
with previous general results by \cite{hol99} and others.

Thus, our results of the retrograde simulations indicate a high probability
for stability within this type of configuration.  Moreover, this
configuration displays regions in the system where bounded chaos can
exist due to the resonances present.  Concerning retrograde orbits,
there is little preference for the planetary starting position.
In summary, we conclude that the $\nu$~Octantis star--planet system
is an interesting case for future observational and theoretical
studies, including additional long-term orbital stability analyses
if the suggested planet is confirmed by follow-up observations.

\section*{Acknowledgments}

This work has been supported by the U.S. Department of Education under
GAANN Grant No. P200A090284 (B.~Q.), the SETI institute (M.~C.) and
the Alexander von Humboldt Foundation (Z.~E.~M.).

\label{lastpage}

\end{document}